\documentclass[apj]{emulateapj}
\usepackage{float,epsfig}
\usepackage{pstricks}
\usepackage{lscape}

\newcommand{\etal}{\mbox{et~al.}}

\def\deg      {{\ifmmode^\circ\else$^\circ$\fi}} 

 \slugcomment{ApJS 2007, COSMOS Special Issue}
 
 \shorttitle{X-ray spectral properties of AGN.}
 \shortauthors{Mainieri et al.}
 
\begin{document}
 
 
 \title{The XMM-{\it Newton} wide-field survey in the COSMOS field.\\ IV: X-ray spectral properties of Active Galactic Nuclei.}
 

%
%
%
 \author{ 
V. Mainieri\altaffilmark{1,2},
G. Hasinger\altaffilmark{1},
N. Cappelluti\altaffilmark{1},
M. Brusa\altaffilmark{1},
H. Brunner\altaffilmark{1},
F. Civano\altaffilmark{3},
A. Comastri\altaffilmark{3},
M. Elvis\altaffilmark{4},
A. Finoguenov\altaffilmark{1},
F. Fiore\altaffilmark{5},
R. Gilli\altaffilmark{3},
I. Lehmann\altaffilmark{1},
J. Silverman\altaffilmark{1},
L. Tasca\altaffilmark{6},
C. Vignali\altaffilmark{3},
G. Zamorani\altaffilmark{3},
E. Schinnerer\altaffilmark{7},
C. Impey\altaffilmark{8},
J. Trump\altaffilmark{8},
S. Lilly\altaffilmark{9},
C. Maier\altaffilmark{9},
R. E. Griffiths\altaffilmark{10},
T. Miyaji\altaffilmark{10},
P. Capak\altaffilmark{11},
A. Koekemoer\altaffilmark{12},
N. Scoville\altaffilmark{11,13},
P. Shopbell\altaffilmark{11},
Y. Taniguchi\altaffilmark{14}}

\altaffiltext{1}{Max Planck Institut f\"ur extraterrestrische Physik, 
       Giessenbachstrasse 1, D--85748 Garching, Germany} 
\altaffiltext{2}{European Southern Observatory, Karl-Schwarzschild-Strasse 2, D--85748 Garching, Germany} 
\altaffiltext{3}{INAF-Osservatorio Astronomico di Bologna, via Ranzani 1, I--40127 Bologna, Italy}
\altaffiltext{4}{Harvard-Smithsonian Center for Astrophysics, 60 Garden Street, Cambridge, MA 02138}
\altaffiltext{5}{INAF-Osservatorio Astronomico di Roma, via Frascati 33, I-00040 Monteporzio Catone (Roma), Italy}

\altaffiltext{6}{Laboratoire d'Astrophysique de Marseille, UMR 6110 CNRS-Universit\'e de Provence, BP8, 13376 Marseille Cedex 12, France}
\altaffiltext{7}{Max Planck Institut f\"ur Astronomie, K\"onigstuhl 17, Heidelberg, D-69117, Germany}
\altaffiltext{8}{Steward Observatory, University of Arizona, 933 North Cherry Avenue, Tucson, AZ 85721}
\altaffiltext{9}{Department of Physics, ETH Zurich, CH-8093 Zurich, Switzerland}
\altaffiltext{10}{Department of Physics, Carnegie Mellon University, 5000 Forbes Avenue, Pittsburgh, PA 15213}
\altaffiltext{11}{California Institute of Technology, MC 105-24, 1200 East
California Boulevard, Pasadena, CA 91125}
\altaffiltext{12}{Space Telescope Science Institute, 3700 SanMartin
Drive, Baltimore, MD 21218}
\altaffiltext{13}{Visiting Astronomer, Univ. Hawaii, 2680 Woodlawn Dr., Honolulu, HI, 96822}
\altaffiltext{14}{Physics Department, Graduate School of Science \&
  Engineering, Ehime University, Bunkyo-cho 2-5, Matsuyama, Ehime 790-8577,
  Japan}
%

%
 
\altaffiltext{$\star$}{Based on observations obtained with XMM-{\it Newton}, an ESA science mission with instruments and contributions directly funded by ESA Member States and NASA}

  
 \begin{abstract} 
 
 We present a detailed spectral analysis of point-like X-ray sources
 in the XMM-COSMOS field. Our sample of 135 sources only includes
 those that have more than 100 net counts in the 0.3-10 keV energy
 band and have been identified through optical spectroscopy. The
 majority of the sources are well described by a simple power-law
 model with either no absorption ($76\%$) or a significant intrinsic,
 absorbing column ($20\%$). The remaining $\sim 4\%$ of the sources
 require a more complex modeling by incorporating additional
 components to the power-law. For sources with more than 180 net
 counts (bright sample), we allowed both the photon spectral
 index $\Gamma$ and the equivalent hydrogen column N$_{\rm H}$
 to be free parameters.  For fainter sources, we fix $\Gamma$ to the
 average value and allow N$_{\rm H}$ to vary. The mean spectral index
 of the 82 sources in the bright sample is $<\Gamma>=2.06\pm0.08$,
 with an intrinsic dispersion of $\sim 0.24$.  Each of these sources
 have fractional errors on the value of $\Gamma$ below $20\%$. As
 expected, the distribution of intrinsic absorbing column densities is
 markedly different between AGN with or without broad optical emission
 lines. We find within our sample four Type-2 QSOs candidates (L$_X >
 10^{44}$ erg s$^{-1}$, N$_H > 10^{22}$ cm$^{-2}$), with a spectral
 energy distribution well reproduced by a composite Seyfert-2
 spectrum, that demonstrates the strength of the wide field XMM/COSMOS
 survey to detect these rare and underrepresented sources. In
 addition, we have identified a Compton-thick (N$_H > 1.5 \times
 10^{24}$ cm$^{-2}$) AGN at z=0.1248. Its X-ray spectrum is well
 fitted by a pure reflection model and a significant Fe K$\alpha$ line
 at rest-frame energy of 6.4 keV.
\end{abstract}
 
 


 
\section{Introduction}

Deep pencil-beam surveys with {\it ROSAT} \citep{hasinger98}, {\it
  Chandra} \citep{brandt01,rosati02,cowie02,alexander03} and XMM-{\it
  Newton} \citep{hasinger01,loaring05} have proved that the majority
of the X-ray background (XRB) is generated by Active Galactic Nuclei
(AGN) both in the soft (0.5-2 keV) and hard (2-10 keV) band. At fluxes
below $\sim 10^{-14}$ erg cm$^{-2}$ s$^{-1}$ in the hard band, the
X-ray source population in these surveys is mainly composed of
obscured AGN.  This supports the suggestion by \cite{setti89} that the
spectral shape of the XRB is due to the integrated contribution of AGN
affected by photoelectric obscuration with a wide range of gas column
density (N$_{\rm H}$) and redshifts.  Since the resolved fraction of
the XRB drops from $\approx 80-90\%$ at $2-6$ keV to $50-70\%$ at
$6-10$ keV \citep{worsley04}, a sizable number of strongly absorbed
AGN may still be missing in the X-ray surveys. An alternative method
to detect heavily absorbed AGN is to select objects that have mid-IR
and radio emission typical of AGN though faint near-IR and optical
fluxes \citep{martinez05}. While this kind of study cannot quantify
which fraction of these mid-IR selected, absorbed AGN would be
detected by X-ray selection, the COSMOS survey \citep{scoville07} will
be able to answer this question due to its rich multi-wavelength
coverage (from radio to X-ray) on a large area of the sky (2
deg$^2$).\\ The XMM-{\it Newton} wide-field survey in the COSMOS field
(XMM-COSMOS, \citealt{hasinger07}), with an unprecedented combination
of wide area coverage and high sensitivity, is providing a large
number of AGN with enough counts to perform a detailed study of their
X-ray spectra. This spectral information, particularly the N$_{\rm H}$
distribution, is a fundamental input parameter to model the XRB
(e.g. \citealt{comastri95,gilli01}). While we anticipate the
completion of the multi-wavelength campaigns including the optical
spectroscopic follow-up within the next few years, we report in this
paper the X-ray spectral fitting results for a preliminary sample of
spectroscopically-identified X-ray sources. The paper is structured as
follows: in \S2 we describe the sample selected on the basis of counts
statistics and optical identification; in \S3 we describe our X-ray
spectral extraction procedure, in \S4 we present the results of the
X-ray spectral analysis, in \S5 we discuss the properties of four
Type-2 QSOs, in \S6 we compare the X-ray and optical classification,
and finally we summarize our conclusions in \S7.\\ Throughout the
paper we assume $H_0=70$ km s$^{-1}$ Mpc$^{-1}$, $\Omega_m$=0.3 and
$\Omega_{\Lambda}$=0.7.

\begin{figure}
\epsscale{1.}
\plotone{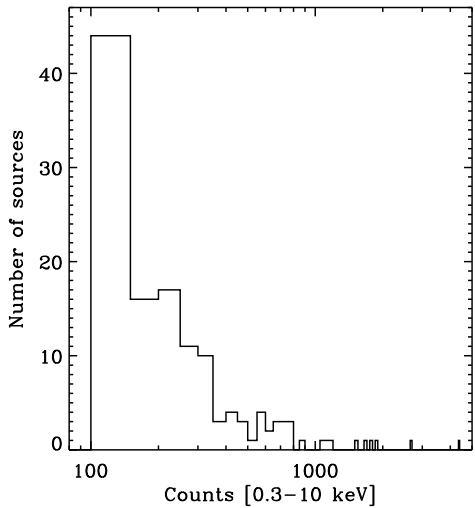}
\caption{Net [0.3-10] keV {\it pn} counts distribution for the sample of 135 X-ray sources used in this work.}
\label{cts_histo}
\end{figure}

\section{Sample selection}
\label{sample}

\begin{figure}
\epsscale{1.}
\plotone{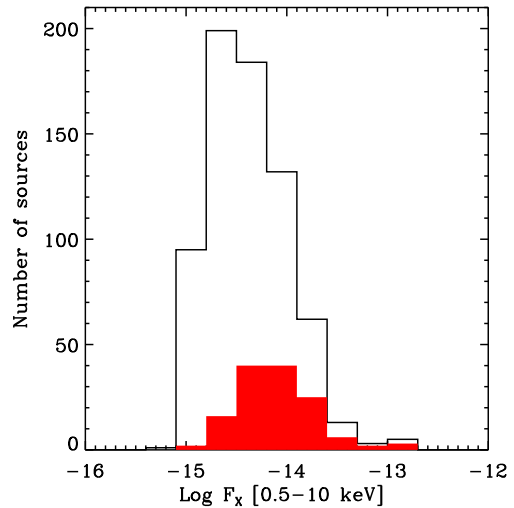}
\caption{X-ray [0.5-10 keV] flux distribution for all the X-ray
  sources (empty histogram) and for the sample of spectroscopically
  identified sources (filled histogram), with more than 100 net
  counts, we analyze in this work.}
\label{histo_fx}
\end{figure}

XMM-$Newton$ has imaged the full 2 deg$^2$ of the COSMOS area down to
the following flux limits in the respective energy bands: $7\times
10^{-16}$ erg cm$^{-2}$ s$^{-1}$ [0.5-2 keV], $4.0 \times 10^{-15}$
erg cm$^{-2}$ s$^{-1}$ [2-10 keV] and $1.0 \times 10^{-14}$ erg
cm$^{-2}$ s$^{-1}$ [5-10 keV] (see Fig. 7 of \citealt{cappelluti07}
for details on the sky coverage as a function of the X-ray flux).  A
general outline of the survey can be found in
\cite{hasinger07}. Further details such as the point-source detection
method and sky area coverage as a function of the X-ray flux are
presented in \cite{cappelluti07}. Our sample is based on the X-ray
catalogue of 1390 point-like sources \citep{cappelluti07}. We limit
our analysis to the sources detected with the EPIC pn$-$CCD ({\it
    pn}) camera \citep{struder01}, in the first 12 XMM-COSMOS
observations since optical spectroscopic follow-up
\citep{trump07,lilly07} has been concentrated in this area ($\sim 1.3$
deg$^2$). These 12 fields are flagged in Table 1 of
\cite{hasinger07}. Reliable optical counterparts
\citep{brusa07} have been determined for $\sim 90\%$ of the sources in
these 12 fields.  We exclude 20 of the 715 X-ray sources in this area
that are classified as ``extended'' from the detection algorithm. The
observed X--ray emission from these sources is likely to be due to a
group or cluster of galaxies, while here we are interested in
selecting AGN. From the remaining 695 X--ray sources, we select
sources with greater than 100 net counts in the [0.3-10] keV energy
band and optical spectroscopic identification.  We further remove one
source that has been identified as a star \citep{trump07}. The final
sample comprises 135 objects.  We show the distribution of their net
counts in the [0.3-10] keV band in Figure \ref{cts_histo} and the
[0.5-10] keV flux distribution (Figure \ref{histo_fx}) that covers a
range of $1.4\times 10^{-15}$, $1.2 \times 10^{-13}$ erg cm$^{-2}$
s$^{-1}$. From their optical spectra, we can further subdivide our
sample based on the presence of broad emission lines: {\it 'Broad Line
AGN'} (BL AGN, 86 objects; FWHM$>$2000 km s$^{-1}$), {\it 'NON-Broad
Line AGN'} (NON-BLAGN, 49 objects; FWHM$<$2000 km s$^{-1}$).  We note
that in this latter class there are objects showing clear signs of
nuclear activity such as high excitation emission lines, as well as
sources with normal galaxy spectra. We compare this purely optical
classification with the X-ray properties of our sources in Sec.
\ref{classification}.

\begin{figure*}
\epsscale{1.}
\includegraphics[angle=0,width= 8.5cm]{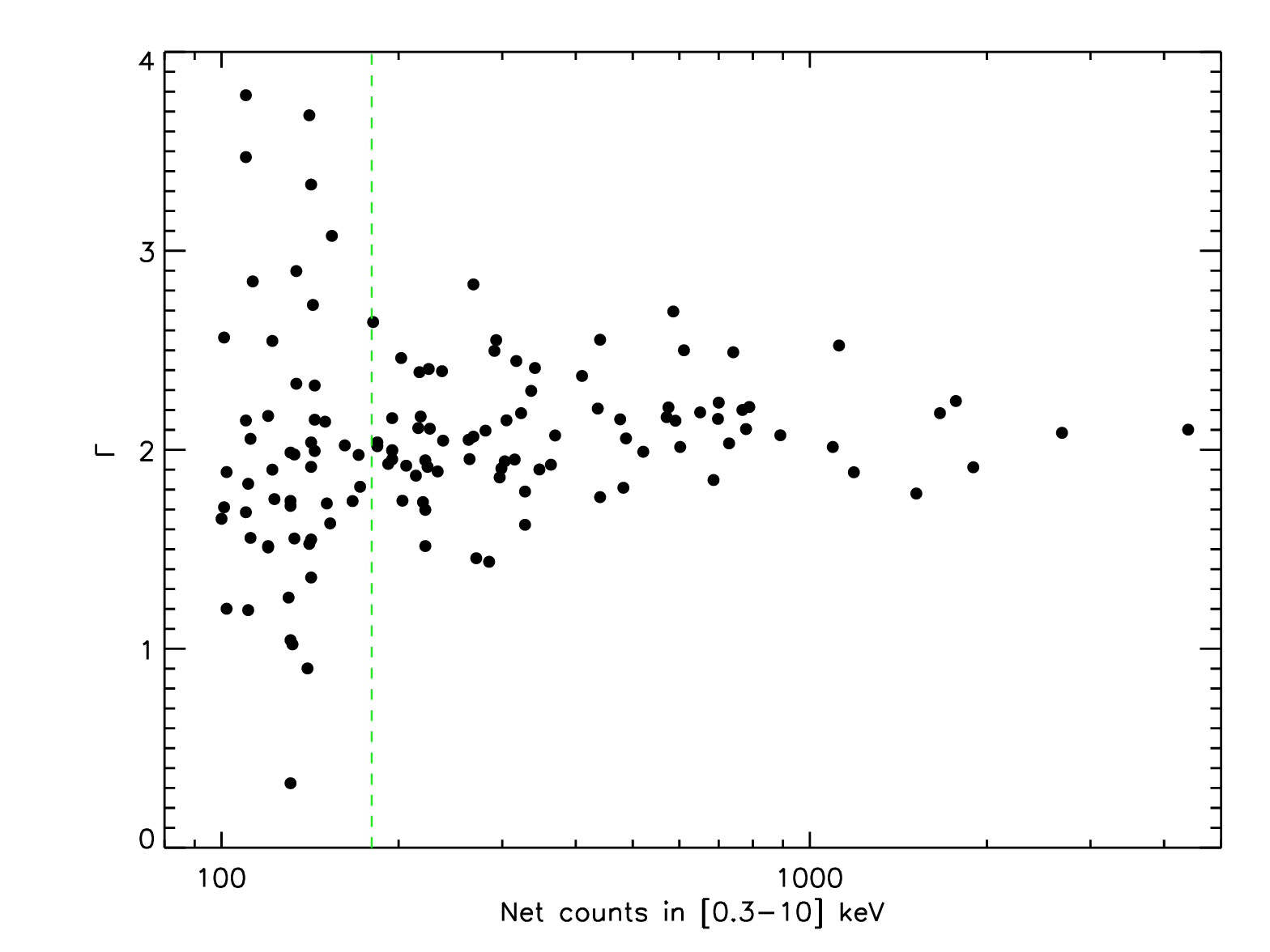}
\includegraphics[angle=0,width= 8.5cm]{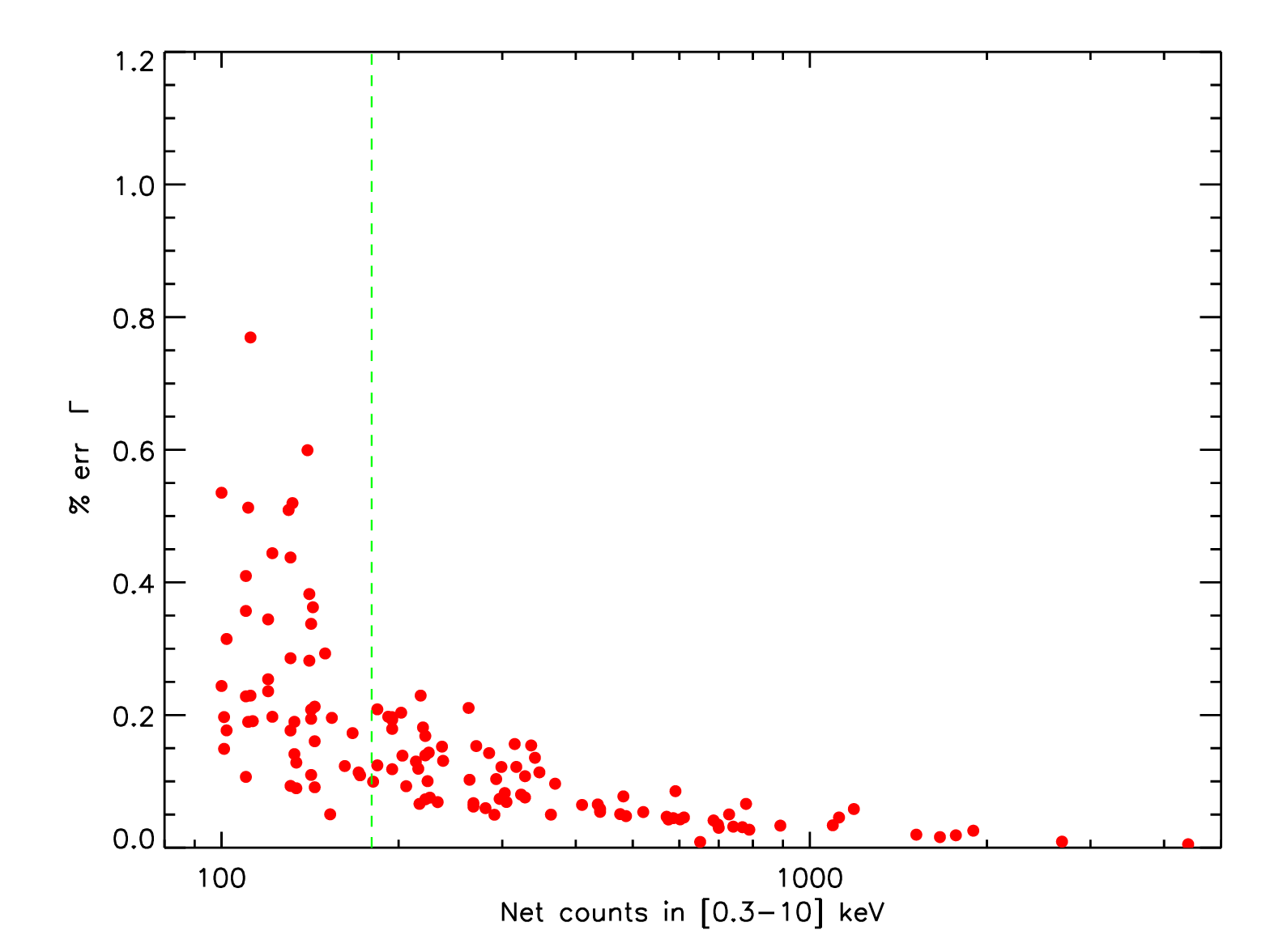}
\caption{{\it Left}: spectral slope value ($\Gamma$) from the fit of
  each single source using the PL model as a function of the net
  counts in the full [0.3-10] keV band. {\it Right}: fractional
  statistical error ($1\sigma$) on $\Gamma$ as a function of the net
  counts in the [0.3-10] keV energy band. The dashed line in both
  plots is the threshold of 180 net counts in the [0.3-10] keV band
  that divides the {\it bright} from the {\it faint} sample.}
\label{gamma_cts}
\end{figure*}

\section{Extraction of X-ray spectral products}

We have implemented an automated procedure to produce the X-ray
spectrum for each source by combining counts from individual
exposures. We have used the latest release of the XMM-{\it Newton}
Science Analysis System
(SAS)\footnote{http://xmm.vilspa.esa.es/external/xmm\_sw\_cal/sas\_frame.shtml}
software package (v 7.0). The task {\it region} has been used to
generate the source and background extraction regions. The source
region is defined as a circle with radius r$_s$ that varies according
to the signal-to-noise and the off-axis angle of the detection to
optimize the quality of the final spectrum. The radii of these regions
are reduced by the task to avoid overlapping with the extraction
regions of nearby sources. All source regions are further excised from
the area used for the background measurement. The task {\it especget}
has been used to extract from the event file the source and background
spectra for each object. The same task generates the calibration
matrices (i.e. arf and rmf) for each spectrum and determines the size
of the source and background areas while updating the keyword BACKSCAL
in the header of the spectra appropriately\footnote{The header keyword
  BACKSCAL is set to 1 for the source spectrum while for the
  background spectrum it is fixed to the ratio between the background
  to source areas.}. The single pointing spectra have been combined
with {\it mathpha} to generate the spectrum of the whole
observation.\footnote{We note that all the XMM-Newton observations in
  the COSMOS field have been performed with the thin filter for the
  {\it pn} camera.} For each source in our sample, we use all the
available counts from the XMM-COSMOS observations, including those
coming from overlapping fields not included in the 12 fields list (see
Fig.1 \citealt{hasinger07}). Finally, in order to use the $\chi ^2$
minimization technique, we bin the spectra with {\it grppha} so that
each bin has at least 20 counts.

\section{Spectral analysis}
\label{analysis}

 \begin{deluxetable*}{rccccccccc}
 \tabletypesize{\scriptsize}
 \tablecaption{Parameters of the best fit model for sources with soft excess
 \label{softexc}}
 \tablewidth{0pt}
 \tablehead{
 \colhead{XID} & \colhead{counts\tablenotemark{a}} & \colhead{Model\tablenotemark{b}} & \colhead{$\chi ^2/d.o.f.$}  & \colhead{$\Gamma$\tablenotemark{c}} & \colhead{N$_{\rm H}$\tablenotemark{d}} &  \colhead{kT\tablenotemark{e} } & \colhead{$\Gamma$\tablenotemark{f}} & \colhead{Redshift} & \colhead{Opt. class\tablenotemark{g}} } 
 \startdata
41 & 315 & APL+po & 0.94 & 1.72$^{2.57}_{1.38}$ & 21.38$^{21.59}_{21.02}$ &  & 2.0 & 0.114 & NLAGN \\
 &  & APL+bb & 0.83 & 1.95$^{2.37}_{1.65}$ & 21.51$^{21.55}_{20.42}$ & 30$^{+2}_{-3}$ & &  & \\ 
106 & 141 & APL+po & 0.25 & 2.0 & 22.33$^{22.66}_{21.98}$ & & 2.0 & 0.710 & gal \\
 &  & APL+bb & 0.26 & 2.0 & 22.28$^{23.00}_{21.96}$ & 121$^{+914}_{-71}$  &  &  &  \\
117 & 111 & APL+po & 0.69 & 2.0 & 22.76$^{23.14}_{22.37}$ &  & 2.0 & 0.936 & gal \\
 &  & APL+bb & 0.59 & 2.0 & 22.57$^{22.86}_{20.42}$ & 81$^{+48}_{-18}$ & &  &  \\
274 & 112  & APL+po & 0.26 & 2.0 & 22.67$^{23.00}_{22.18}$ & & 2.0 & 0.677 & gal  \\
 &  & APL+bb & 0.24 & 2.0 & 22.55$^{23.00}_{20.42}$ & 124$^{+878}_{-119}$ & &  &  \\
 \enddata
 \tablenotetext{a}{Net {\it pn} counts in the [0.3-10] keV energy range.}
 \tablenotetext{b}{Best fit model: APL+po=absorbed power-law plus an extra power-law for the soft excess; APL+bb=absorbed power-law plus a blackbody for the soft excess.}
 \tablenotetext{c}{Slope of the powerlaw model (photon index).}
 \tablenotetext{d}{ ${\rm Logarithm}$ of the intrinsic absorption (cm$^{-2}$).}
 \tablenotetext{e}{Temperature (in eV) of the blackbody used to model the soft excess.}
 \tablenotetext{f}{Slope of the extra power-law used to model the soft excess.}
 \tablenotetext{g}{Optical classification, see Sec. \ref{classification} for details.}
  \end{deluxetable*}

\cite{tozzi06} have shown by extensive simulations that below 50
counts the best fit values obtained using Cash statistics
\citep{cash79} are more accurate than those obtained with the $\chi
^2$.  For greater than 50 counts, the two methods give equivalent
results. Since we limit our analysis in this paper to sources with
more than 100 counts, we are confident that the results obtained with
the $\chi ^2$ minimization technique are accurate. We use
XSPEC\footnote{http://heasarc.gsfc.nasa.gov/docs/xanadu/xspec/}
(v11.3.2p) for our spectral fitting analysis. We first fit the data
with two basic input models: a simple {\it powerlaw} ({\it PL}) and a
{\it powerlaw} modified by intrinsic absorption at the redshift of the
source ({\it APL}). Both models include an additional component to
account for photoelectric absorption due to the Galactic column
density that is fixed to the value in the COSMOS region (N$_{\rm
H}^{\rm Gal}\sim 2.7
\times 10^{20}$ cm$^{-2}$, \citealt{dickey90})\footnote{This is an
average value for the Galactic N$_{\rm H}$ in the COSMOS area where
N$_{\rm H}^{\rm Gal}$ is in the range [2.5-2.9] $\times 10^{20}$
cm$^{-2}$.  This range in Galactic column density does not affect the
results of our spectral analysis.}. The PL model is made of two
XSPEC components {\it wabs*zpowerlw}, while the APL model consists of
the combination of three different components {\it
wabs*zwabs*zpowerlw}. The {\it wabs} model describes the photoelectric
absorption using Wisconsin cross-sections \citep{morrison83} and its
only parameter is the equivalent hydrogen column density ({\it zwabs}
has the redshift as an additional parameter). {\it zpowerlw} is a
simple power law parameterized by the photon index, the redshift and a
normalization factor.\footnote{ We refer the reader to
http://heasarc.gsfc.nasa.gov/docs/xanadu/xspec/ for further details on
the spectral models.} The model fits yield the power-law photon index
$\Gamma$, the X-ray luminosity in the [0.5-2] and [2-10] keV
rest-frame bands, and from the {\it APL} model also the intrinsic
column density ${\rm N_H}$ . We notice that the dispersion of $\Gamma$
for our sample increases significantly for sources with low counts
statistics (Fig. \ref{gamma_cts}, left panel) and in particular the
fractional error becomes quite large (Fig.
\ref{gamma_cts}, right panel). Above 180 net counts, the fractional
error remains below $20\%$. Hence, we split our sample in two: {\it
sample-1} including 82 sources with more than 180 net counts, and {\it
sample-2} having 53 sources with less than 180 counts. For {\it
sample-1}, we allow both $\Gamma$ and ${\rm N_H}$ free to vary, while
we fix $\Gamma$ to the average value, obtained with {\it sample-1},
for lower count sources ({\it sample-2}).\\ For all the 135 X-ray
sources, we perform a spectral fit using both {\it PL} and {\it APL}
models. We label a source as X-ray absorbed in those cases for which
the {\it APL} model is a better fit, than the pure {\it PL}, with a
confidence level threshold of $90\%$ based on an F-test.\\  The
output of our spectral analysis is reported in
Tab. \ref{xspec_catalogue}. The table has the following structure: IAU
name (col.1), identification number (xid, col.2), X-ray coordinates
(col.3-4), net detected X-ray counts in the [0.3-10] keV band
(col. 5), spectroscopic redshift (col. 6), best fit model (col. 7),
spectral index $\Gamma$ (col. 8), intrinsic column density N$_{\rm H}$
(col. 9), X-ray fluxes (col. 10-11-12), de-absorbed X-ray luminosities
(col. 13-14-15).


\subsection{Notes on some individual sources}

\begin{figure*}
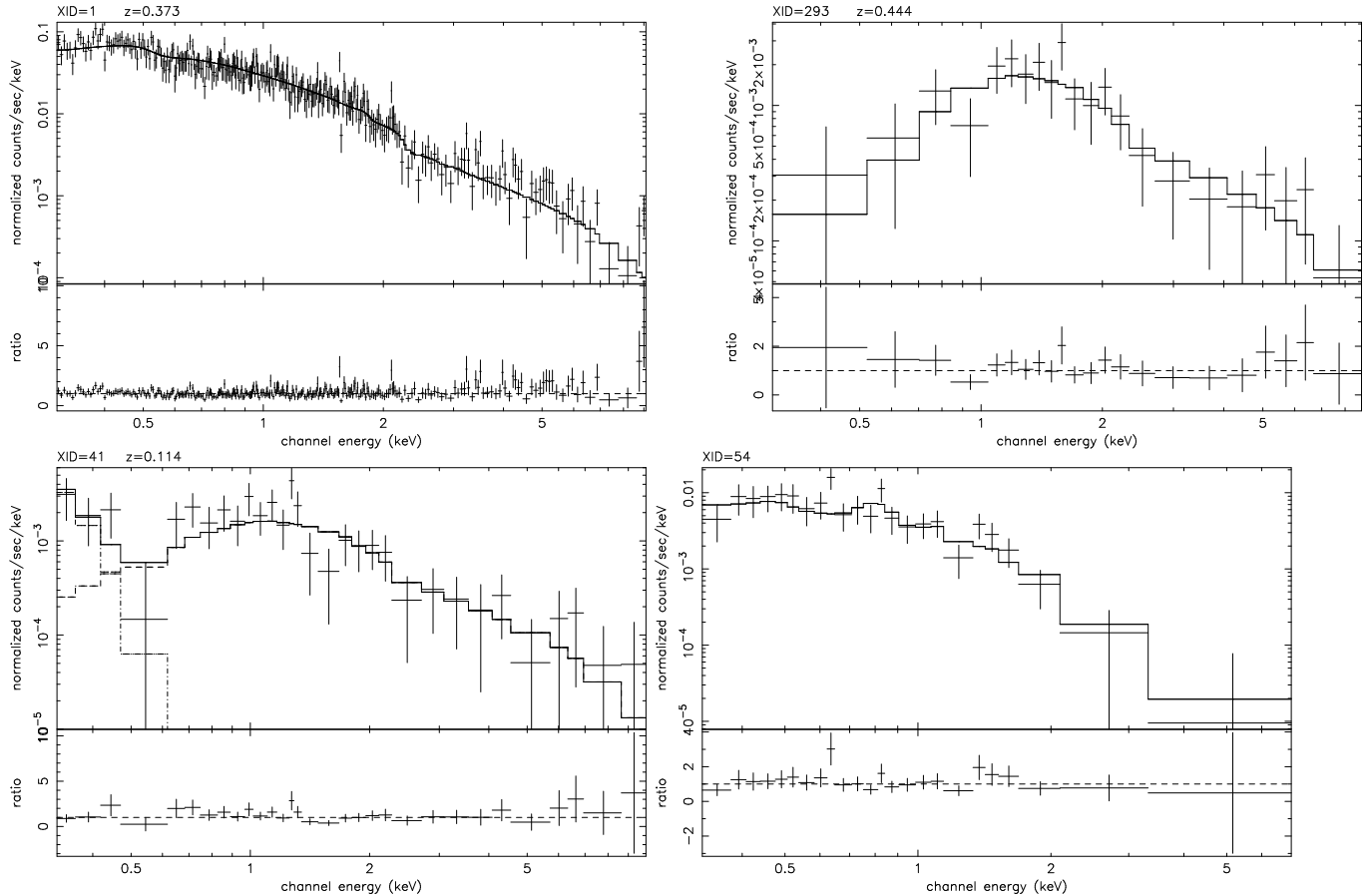

\includegraphics[angle=-90,width= 8.5cm]{f4a.eps}
\includegraphics[angle=-90,width= 8.5cm]{f4b.eps}
\includegraphics[angle=-90,width= 8.5cm]{f4c.eps}
\includegraphics[angle=-90,width= 8.5cm]{f4d.eps}
\caption{Examples of X-ray spectra with different best fit model. {\it
    Top-left}: unabsorbed power-law (PL); {\it top-right}: absorbed
  power-law (APL); {\it bottom-left}: absorbed power-law plus a
  black-body component to model the soft excess; {\it bottom-right}:
  thermal emission parameterized with a Raymond-Smith model.}
\label{spec_example}
\end{figure*}

For each source we carefully check the results obtained with the basic
{\it PL} and {\it APL} models and, if significant residuals are
present, we refine the fit using more complex models. We show in Fig.
\ref{spec_example} a representative X-ray spectrum for each one of the
different best-fit models. We use the F-test and a confidence level
threshold of $90\%$ to choose between the different models.

\subsubsection{Soft excess}

A clear {\it soft excess} is present in four of our sources (xid 41,
106, 117, 274). This feature, first observed with EXOSAT
\citep{arnaud85,turner89}, has been confirmed by XMM-{\it Newton}
observations (e.g. \citealt{pounds02,porquet04,gallo06}), but its
origin is still uncertain. Such a soft component may be the high
energy tail of the UV bump (a blackbody model is appropriate in this
case), or can be due to reprocessed emission scattered along our line
of sight by a photo-ionized gas located just above the obscuring torus
(an additional power-law with the spectral index fixed to the value of
the hard X-ray primary power-law is a good parameterization of this
scenario). We fit these four sources adding to the basic APL model an
extra component represented either by a power-law ({\it po}) or a
blackbody ({\it bb}\footnote{This is a blackbody spectrum defined
by the temperature kT in keV and a normalization factor.}) according
to the two physical scenarios mentioned above. We report in Tab.
\ref{softexc} the parameters of the additional component in the fit of
these four sources. We are not able to distinguish on a statistical
basis between the two models given the similar values of
$\chi^2$. Nevertheless, we notice that all these four sources present
intrinsic absorption and therefore we exclude that the soft-excess of
these objects is due to the high energy tail of the UV bump (APL+bb).

 \begin{deluxetable}{rcccc}
 \tabletypesize{\scriptsize}
 \tablecaption{Parameters of the {\it gauss} additional component for the sources with Fe line
 \label{Feline}}
 \tablewidth{0pt}
 \tablehead{
 \colhead{XID} &  \colhead{$\sigma$\tablenotemark{a} } & \colhead{EW\tablenotemark{b} }& \colhead{Redshift} & \colhead{Opt. class\tablenotemark{c}} } 
 \startdata

2028 & 616$^{+364}_{-224}$ & 2754$^{+1628}_{-1002}$ & 0.784 & gal \\ 
2043 & 179$^{+120}_{-115}$ & 748$^{+502}_{-481}$  & 0.668 & gal \\ 
2608 & $281^{+408}_{-175}$ &  $792^{+1151}_{-493}$  & 0.125 & gal \\ 
 \enddata
 \tablenotetext{a}{Observed width of the line in eV.}
 \tablenotetext{b}{Rest frame equivalent width of the line in eV}
 \tablenotetext{c}{Optical classification, see Sec. \ref{classification} for details.}

  \end{deluxetable}

\subsubsection{Fe K$\alpha$ line}

\begin{figure}
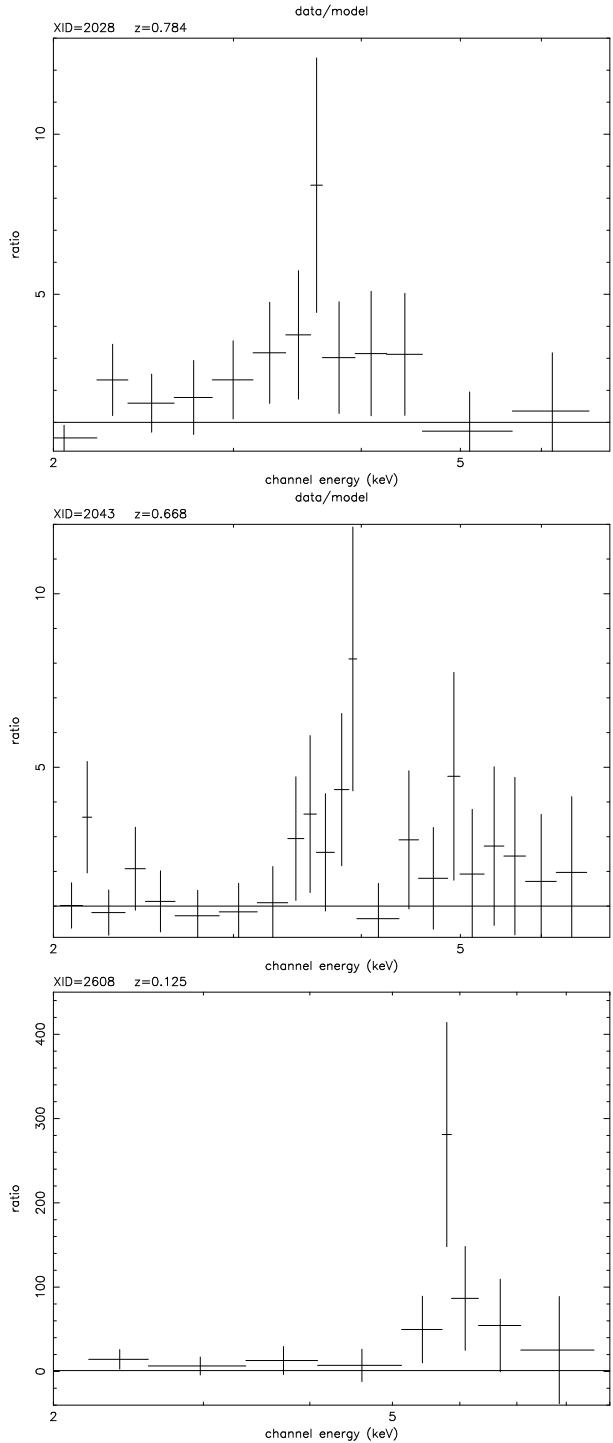

\includegraphics[angle=-90,width= 8cm]{f5a.eps}\\
\includegraphics[angle=-90,width= 8cm]{f5b.eps}\\
\includegraphics[angle=-90,width= 8cm]{f5c.eps}
\caption{The ratio of the data versus a powerlaw model (top panels) or a pexrav model (bottom figure) around the energy of the Fe K$\alpha$ line for the three sources with significant detection of this feature. }
\label{Feratio}
\end{figure}

Three sources show significant features ascribable to the redshifted
Fe K$\alpha$ emission line: xid 2028, 2043, 2608. For these sources we
add a Gaussian component ({\it gauss}) to the model, fixing the line
energy to $6.4/(1+z)$ keV. The best fit values of interesting
parameters are reported in Tab. \ref{Feline}. We show in Fig.
\ref{Feratio} the ratio of the data versus the model (powerlaw for xid
2028, 2043 and pexrav\footnote{An exponentially cut off power law
spectrum reflected from neutral material. We refer the reader to
\cite{magdziarz95} for a detailed description of such model.} for xid 2608) in an energy range around the
expected location of the Fe K$\alpha$ line.  Interestingly all three
of these sources do not show sign of AGN activity from their optical
spectra and are therefore classified as 'galaxy'.

\subsubsection{Thermal emission?}

\begin{figure}
  \epsscale{1.}
\plotone{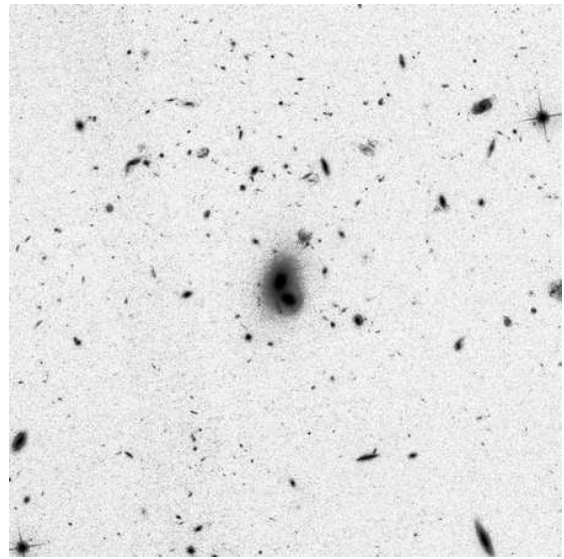}
\caption{ACS image of source xid 54. The cutout is 90 arcsec on a
  side.}
\label{acs54}
\end{figure}

Source xid 54, if fitted with an APL model, gives a large value for
the spectral slope ($\Gamma > 3$) and significant residuals in the
0.3-10 keV energy range. An alternative description of its spectrum is
obtained assuming we are observing thermal emission, parameterized
with a Raymond-Smith model \citep{raymond77} with a temperature
kT$= 1.6^{+0.4}_{-0.2}$ keV fixing the metalicity to $0.3$
solar. Source xid 54 is identified with two interacting galaxies (see
Fig. \ref{acs54}) at redshift z$=0.350$ with no sign of AGN activity
from its optical spectrum. Its X-ray luminosity of $3 \times 10^{42}$
erg s$^{-1}$ is larger than that expected for early-type galaxies
\citep{matsushita01} and, from the optical imaging, there is a
concentration of galaxies around xid 54 with the same photometric
redshifts therefore supporting the idea that we are looking at the
X-ray emission from a group of galaxies.  Nevertheless, we can not
exclude with the current data that a fraction of the X-ray flux of
source 54 could come from an absorbed nucleus (e.g XBONGS,
\citealt{comastri02}) or from discrete sources like LMXBs or HMXBs in the
galaxy. A Chandra observation with its higher angular resolution could
possibly locate discrete sources inside xid 54.

\subsubsection{XID=2608 : a Compton-thick AGN?}
\label{comptonthick}

\begin{figure*}
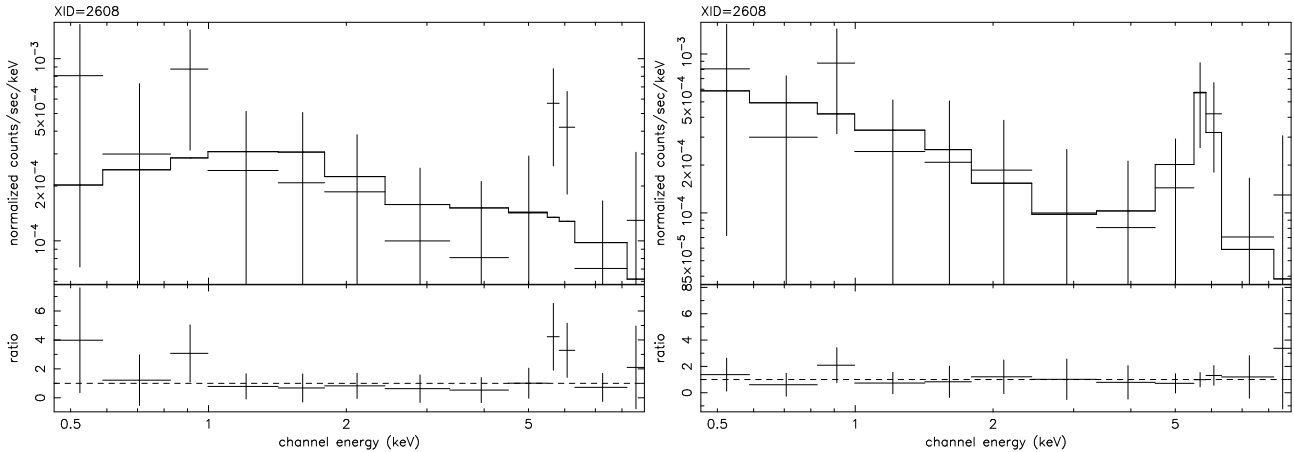

\includegraphics[angle=-90,width= 8.5cm]{f7a.eps}
\includegraphics[angle=-90,width= 8.5cm]{f7b.eps}
\caption{  The X-ray fit of source xid 2608 with  the basic APL model 
  ({\it left}) and a pure reflection model plus a Gaussian
  line ({\it right}).}
\label{2608_fit}
\end{figure*}

An additional source that requires a more complex modeling of its
spectrum is xid 2608. The fit with the APL model gives an extremely
flat value for $\Gamma$ ($\approx 0.3$) and large residuals at both
low and high energies (see left panel of Fig.
\ref{2608_fit}).  \cite{hasinger07} found that this source is located
in an area populated by local Compton-thick Seyfert-2 galaxies in an
X-ray color-color diagram (see Fig. 12 of \citealt{guainazzi05}).
This, together with other evidences based on lines ratios from the
optical spectrum, supports the hypothesis that source 2608 is a
heavily absorbed AGN.

We use the 131 net counts from the pn camera for this source to study
more in detail its X-ray spectrum. A pure reflection component model
({\it pexrav}) is a better description compared to the APL model
according to an F-test with a confidence level of $95\%$. Nevertheless
this fit leaves a clear residual around the expected position of the
6.4 keV Fe K$\alpha$ line. The best-fit model for xid 2608 is a pure
reflection model plus a Gaussian line at 6.4 keV rest-frame ({\it
pexrav + gauss}). The details for the different spectral fits are
reported in Tab. \ref{xid2608}.  The presence of the Fe K$\alpha$
fluorescent line at 6.4 keV is significant at $95\%$ according to an
F-test. The presence of the line is a clear sign that the source is
heavily absorbed, but a useful observable to confirm its Compton-thick
nature is the Equivalent Width (EW) of the same line. The nominal best
fit value for the EW ($792^{+1151}_{-493}$ eV) is higher than the
maximum (600 eV) observed EW in Compton-thin objects
\citep{turner97}. This supports the idea that source 2608 is a
Compton-thick AGN, although we have to mention that with the current
photons statistics, the error for the observed flux of the line ( and
consequently for the EW) is still large. We are confident that an
improved result will come after the completion of the additional 600
ksec XMM observations awarded in AO4.\\ Another diagnostic on the
Compton-thick nature of this source could be the {\it thickness
  parameter} T$=$F[2-10keV]/F[OIII]. A high quality optical spectrum
for this source is available in the Sloan Digital Sky Survey archive
and we obtain a value for F[OIII] from the analysis of
\cite{kauffmann03}.  The [OIII] flux has been corrected for the
extinction toward the narrow-line region as deduced from the Balmer
decrement. We obtain T$=3.8$ which is in a ``grey area'' where both
Compton thick and less absorbed AGN are located (see for example Fig.
1 of \citealt{bassani99}).

In Fig. \ref{lx_nh}, we assume as a lower limit for the column density
of xid 2608 the value $1.5 \times 10^{24}$ cm$^{-2}$ where the Compton
optical depth is equal to unity and the directly transmitted nuclear
emission is strongly suppressed in the [0.3-10] keV band. For the
luminosity of this object, if we assume that only $3\%$ of the flux
has been reflected, we obtain a value of $\sim 7.4 \times 10^{43}$ erg
s$^{-1}$, while for reflected fractions between $10-1\%$ the
luminosity would be in the range $(0.2-2.2) \times 10^{44}$ erg
s$^{-1}$.

 \begin{deluxetable}{cccccc}
 \tabletypesize{\scriptsize}
 \tablecaption{Parameters of the best fit model for source xid 2608
 \label{xid2608}}
 \tablewidth{0pt}
 \tablehead{
 \colhead{Model\tablenotemark{a}} &  \colhead{$\Gamma$} & \colhead{N$_{\rm H}$\tablenotemark{b}} & \colhead{EW\tablenotemark{c}} & \colhead{$\chi^2$} & \colhead{d.o.f.} } 
 \startdata
APL & 2.0 & $0.16^{+0.75}_{-0.16}$ &  & 9.3 & 11 \\ 
pexrav & 2.0 &  &  & 4.1 & 9 \\ 
pexrav+gauss & 2.0 & & $792^{+1151}_{-493}$  & 1.7 & 7 \\ 
 \enddata
 \tablenotetext{a}{Best fit model: {\it APL} = absorbed power-law;
   {\it pexrav} = pure reflection model; {\it pexrav+gauss} = pure
   reflection model plus a Gaussian line.}
 \tablenotetext{b}{Hydrogen column density in unit of $10^{22}$ cm$^{-2}$.}
 \tablenotetext{c}{Equivalent width of the Fe K$\alpha$ line expressed in eV.}
  \end{deluxetable}

\subsection{Spectral properties of the sample}

\begin{figure}
  \epsscale{1.}
\plotone{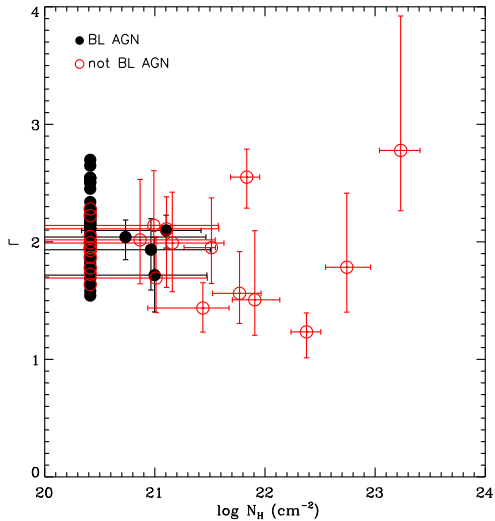}
\caption{$\Gamma$ versus N$_{\rm H}$ for the X-ray sources with more
  than 180 net counts in [0.3-10] keV (bright sample) and
  spectroscopically identified. Filled circles are BL AGN, while empty
  circles are not BL AGN. Error bars correspond to 1$\sigma$. To
  simplify the figure, we did not report the error bars on $\Gamma$
  for unabsorbed sources and plotted them to N$_{\rm H}$=N$^{gal}_{\rm
    H}\approx 2.7 \times 10^{20}$ cm$^{-2}$.}
\label{gamma_nh}
\end{figure}

As mentioned in Sec. \ref{analysis}, we leave both $\Gamma$ and
N$_{\rm H}$ free to vary when fitting the sources in {\it
  sample-1}. The results of this analysis are summarized in Fig.
\ref{gamma_nh}. The average value of $\Gamma$ does not change as a
function of N$_{\rm H}$ as already noticed in deep surveys (i.e.
\citealt{mainieri02}). We obtain, using the weighted mean, $<\Gamma
>=2.06 \pm 0.08$ and the observed dispersion of the distribution of
the best fit values is $\sigma \approx 0.25$. As the typical error in
a single measurement of $\Gamma$ is $\Delta \Gamma = 0.09$, assuming
that both statistical errors and the intrinsic dispersion are
distributed as a Gaussian, the intrinsic scatter in $\Gamma$ is
$\sigma_{int}\sim0.24$. For comparison with X-ray spectral studies in
a similar X-ray flux range of our sample, \cite{mateos05a} from a
large sample of serendipitous sources detected with XMM-{\it Newton}
in a $\sim 3.5$ deg$^2$ area, obtained $<\Gamma>=1.96 \pm 0.01$;
\cite{perola04} in the spectroscopic analysis of the HELLAS2XMM 1df
found $<\Gamma>=1.90 \pm 0.22$; \cite{page06} from the spectral fit of
AGN in the $13^H$ XMM-{\it Newton}/Chandra deep field found
$<\Gamma>=2.0 \pm 0.1$ with an intrinsic dispersion $\sigma \approx
0.36$. All these measurements are consistent with each other within
the uncertainties. Adopting the optical classification described in
Sec.  \ref{sample} the mean value for the spectral slope for BLAGN (58
sources) is $<\Gamma >=2.09$ with a dispersion of $\sigma \approx
0.26$, while for not BLAGN (24 sources) we obtain $<\Gamma >=1.93$ and
$\sigma \approx 0.29$.  Furthermore, we confirm that the average value
of the photon index does not vary with redshift in the range
z=[0.0,3.0] covered by our sample, thus confirming previous findings
(e.g. see Fig. 9 of \cite{piconcelli03} for a compilation from the
literature)

\begin{figure}
  \epsscale{1.} \plotone{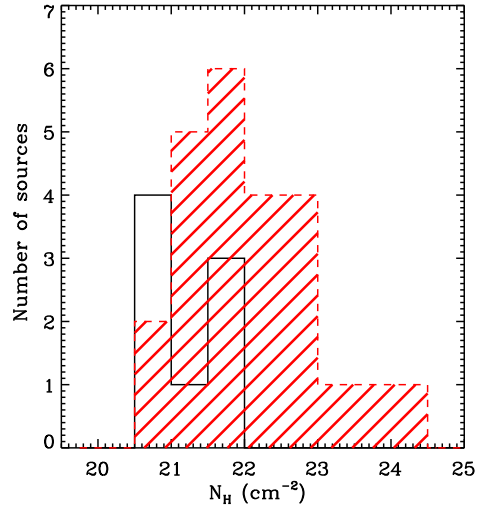}
\caption{Intrinsic column density (N$_{\rm H}$) distribution for BL AGN 
(empty histogram) and NOT BL AGN (hatched histogram) with
intrinsic absorption in excess of the Galactic column density.}
\label{nh_histo}
\end{figure}

\begin{figure}
  \epsscale{1.} \plotone{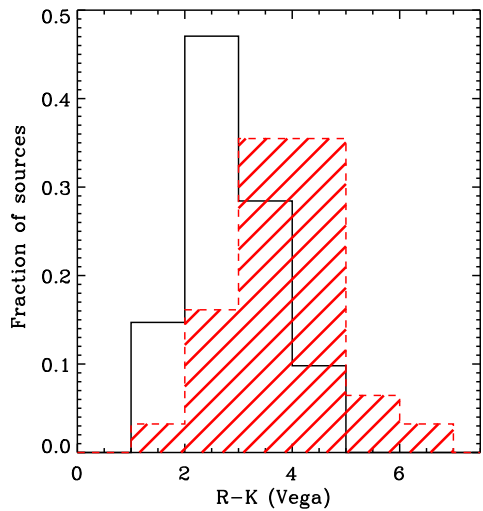}
\caption{R-K colors (Vega) distribution for sources with PL as best fit model 
(empty histogram) and for sources with 'APL' as a best fit model (hatched histogram).}
\label{rk_histo}
\end{figure}

The other physical quantity that we measure from the spectral fitting
is the column density N$_{\rm H}$. In this case, we consider all our
135 sources since N$_{\rm H}$ has been left free to vary both in {\it
sample-1} and {\it sample-2}.  According to an F-test, 32 X-ray
sources do require intrinsic absorption in excess to the Galactic one,
at a confidence level larger than $90\%$. Therefore a fraction as
large as $24\%$ of our sample is made of X-ray absorbed AGN. Fig.
\ref{nh_histo} shows the distribution of N$_{\rm H}$ values for these
sources. We note that the observed N$_{\rm H}$ distribution
refers only to the sources inside the region in the N$_{\rm
H}$-L$_{\rm X}$-z space delimited by the count-rate detection
threshold of our survey. This introduces a bias against absorbed
sources, and therefore the fraction of absorbed sources detected in
our study has to be considered a lower limit. According to the most
recent population synthesis model of the XRB \citep{gilli06}, in the
band used to select our sample the expected fraction of obscured
source with column density N$_{\rm H}>10^{21}$ cm$^{-2}$ is $20\%$
that is consistent with what we found. In Fig. \ref{nh_histo} we
divide BL AGN from NOT BL AGN. The visual impression that NOT BL AGN
have larger column densities than BL AGN is confirmed by a
Kolmogorov-Smirnov test that gives a probability larger than $99.9\%$
that the two distributions are different. No object with N$_{\rm H} >
10^{22}$ cm$^{-2}$ shows broad lines in its optical spectrum.
Nevertheless, $9\%$ (8/86) of the BL AGN in our sample do show some
intrinsic absorption in their X-ray spectra (see also
\citealt{mittaz99,fiore01,page01,schartel01,tozzi01,mainieri02,brusa03,perola04,mateos05}).\\
In Fig \ref{rk_histo} we show the distributions of the R-K (Vega)
colors for the sources with 'PL' as best fit model (empty histogram)
and for the ones that instead require an absorbed power-law (hatched
histogram). The two distributions are significantly different according
to a Kolmogorov-Smirnov test with a probability of $99.99\%$.  The
X-ray sources that require an absorption component in their spectral
fit are on average redder, suggesting a correlation between X-ray
absorption and optical to near-IR colors. On the contrary, sources
that do not show absorption in their X-ray spectra have bluer color
typical of optically selected, unobscured quasars. These results
confirm those obtained from an analysis based on HR values made by
\cite{brusa07} (see their Fig. 10). Nevertheless the interpretation
of this correlation between X-ray absorption and optical to near-IR
colors is not straightforward since we are sampling different scales
in the two measurements (i.e. nucleus with the X-ray data and
nucleus$+$host galaxy with the R-K colors).

\begin{figure}
  \epsscale{1.} \plotone{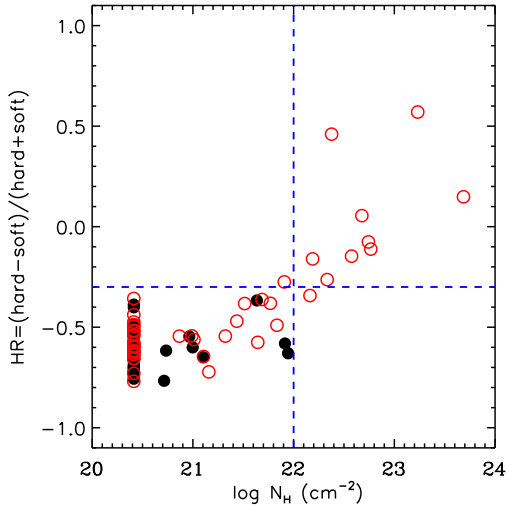}
\caption{HR defined using the [0.5-2] (soft) and [2-10] (hard) bands 
  versus the column density derived from the spectral fitting
  analysis. Only sources with errors on the HR smaller than 0.3 have
  been plotted. Filled circles are BL AGN, while empty circles are
  NOT BL AGN. The horizontal dashed line corresponds to
  HR$=-0.3$ used to separate absorbed and unabsorbed sources, while
  the vertical dashed line indicates a column density equal to
  $10^{22}$ cm$^{-2}$. }
\label{hr_nh}
\end{figure}

When the number of counts in a source is inadequate to perform a
spectral fit, a widely used tool to study the general spectral
properties of an X-ray source is the hardness ratio $HR=(H-S)/(H+S)$,
where {\it H} are the counts in the [2-4.5] keV band and {\it S} those
in the [0.5-2] keV energy band. In Fig. \ref{hr_nh} we show the
hardness ratio values versus the amount of intrinsic absorption
derived from our spectral analysis (both {\it sample-1} and {\it
sample-2}). A clear correlation between the two quantities is present:
$90\%$\footnote{We note that the only source with N$_{\rm H} >
10^{22}$ cm$^{-2}$ and HR$<-0.3$ shows a soft excess in its X-ray
spectrum.} of the sources with N$_{\rm H} > 10^{22}$ cm$^{-2}$ have
HR$>-0.3$ and $99\%$ of the sources with N$_{\rm H} < 10^{22}$
cm$^{-2}$ have HR$<-0.3$.  Therefore, although one has to remember
that the HR is a strong function of redshift (e.g. Fig. 8 in
\citealt{szokoly04}), it is still possible to use HR for statistical
studies.

\begin{figure}
\epsscale{1.}\plotone{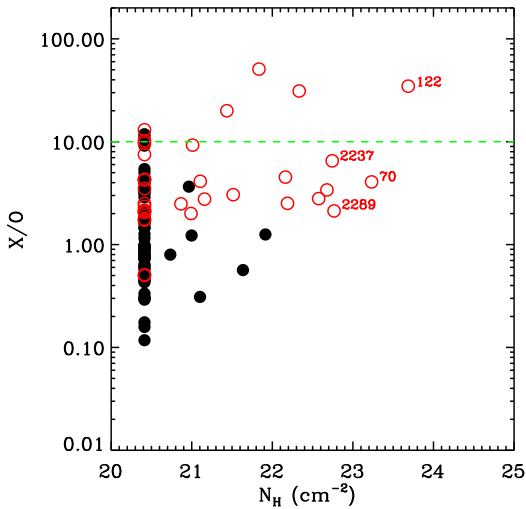}
\caption{X/O (f[2-10 keV]/F(R)) versus N$_{\rm H}$ values. The filled
  circles are BL AGN, the empty circles NOT BL AGN. We label the four
  Type-2 QSOs candidates. The horizontal dashed line indicates the
  value X/O=10. }
\label{nh_lx_xo}
\end{figure}

Another diagnostic that can yield important information on the nature
of X-ray sources is the X-ray-to--optical flux ratio (e.g.,
\citealt{maccacaro88,stocke91}).  The majority of the AGN have
X-ray-to-optical flux ratios (X/O) of $0.1<$X/O$<10$ (e.g.,
\citealt{akiyama00,lehmann01}), but {\it Chandra} and XMM-{\it Newton}
surveys have shown that there is a non negligible population of AGN
with high X/O ($>10$) and that a large fraction of them are obscured,
and possibly high-redshift, Type-2 QSOs (e.g.,
\citealt{fabian00,mainieri02,fiore03,mignoli04,mainieri05}). For
comparison with the literature, we define X/O as the ratio between the
X-ray flux in the [2-10] keV band and the flux in the optical {\it R}
band. In Fig.  \ref{nh_lx_xo} we plot the X/O values for the sources
in our sample versus the N$_{\rm H}$. Out of the seven sources that
have X/O$>10$, four show absorption in their X-ray spectra (APL) and
one is a Type-2 QSO. We notice that the other three Type-2 QSO
candidates in our sample (see Sec.  \ref{qso2}) have X/O values inside
the range $0.1<$X/O$<10$ where most of the optical or soft X-ray
selected AGN are located. Since we limit our analysis to the brighter
X-ray sources and the spectroscopic follow-up is not complete, we
postpone any further analysis on the nature of X/O$>10$ sources to a
future paper.

\section{Type-2 QSO candidates}
\label{qso2}

\begin{figure}
  \epsscale{1.} \plotone{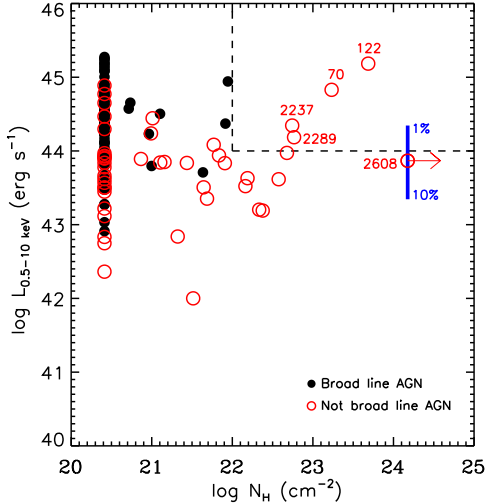}
  \caption{Intrinsic, de-absorbed X-ray luminosity in the [0.5-10] keV
    band vs. N$_{\rm H}$.  The filled symbols are BLAGN, while the
    empty symbols are not BLAGN.  For source xid 2608 we assume a
    lower limit on N$_{\rm H}$ of $1.5 \times 10^{24}$ cm$^{-2}$ and
    for the luminosity we estimate a value of $\sim 7.4 \times 10^{43}$
    erg s$^{-1}$ assuming that a fraction of $3\%$ is reflected (the
    error bar shows the luminosity range covered assuming that the
    reflected fraction is between $1\%$ and $10\%$). See Sec.
    \ref{comptonthick} for details. The dashed lines define the
    ``Type-2 QSO region''.}
\label{lx_nh}
\end{figure}

\begin{figure*}
\epsscale{.60}
\plottwo{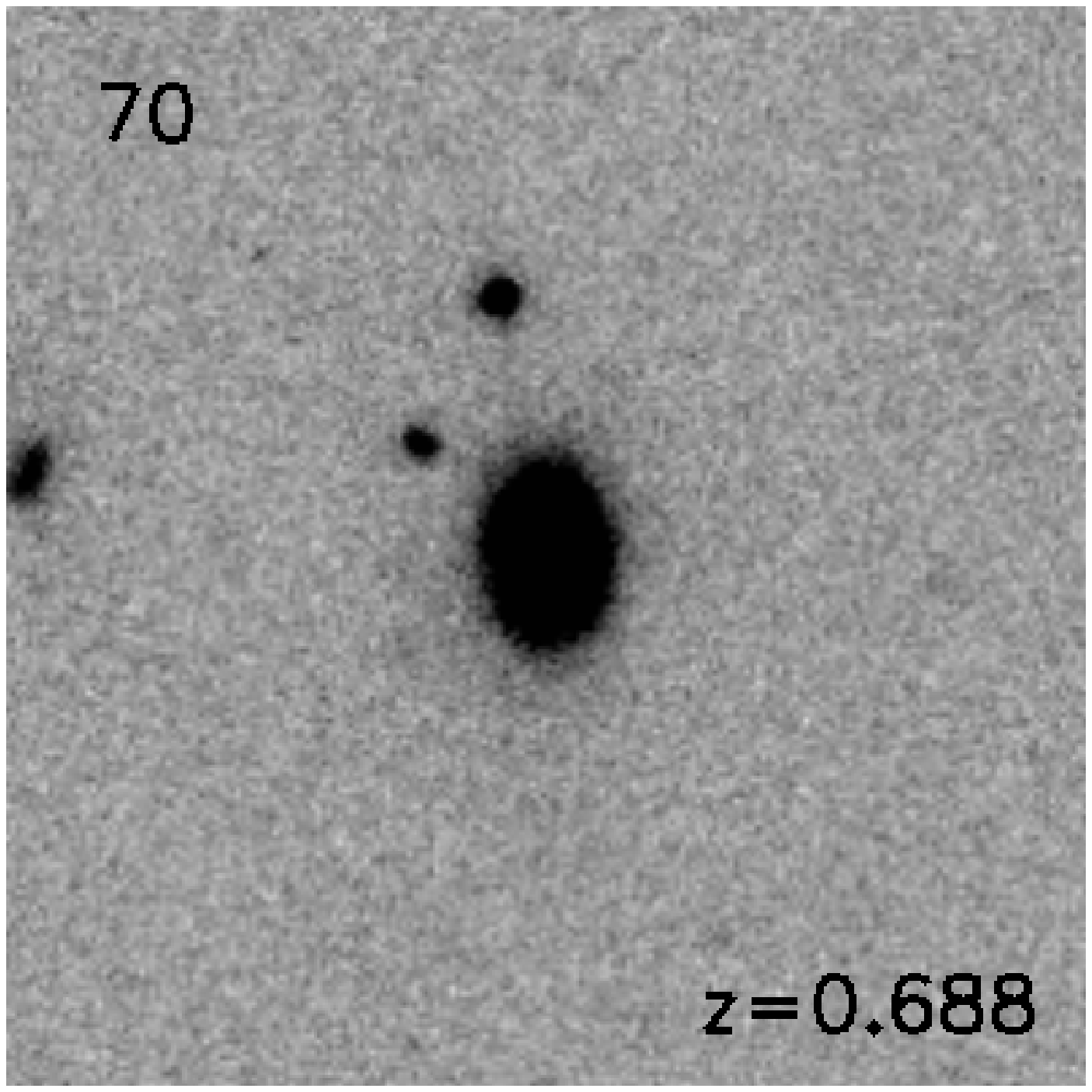}{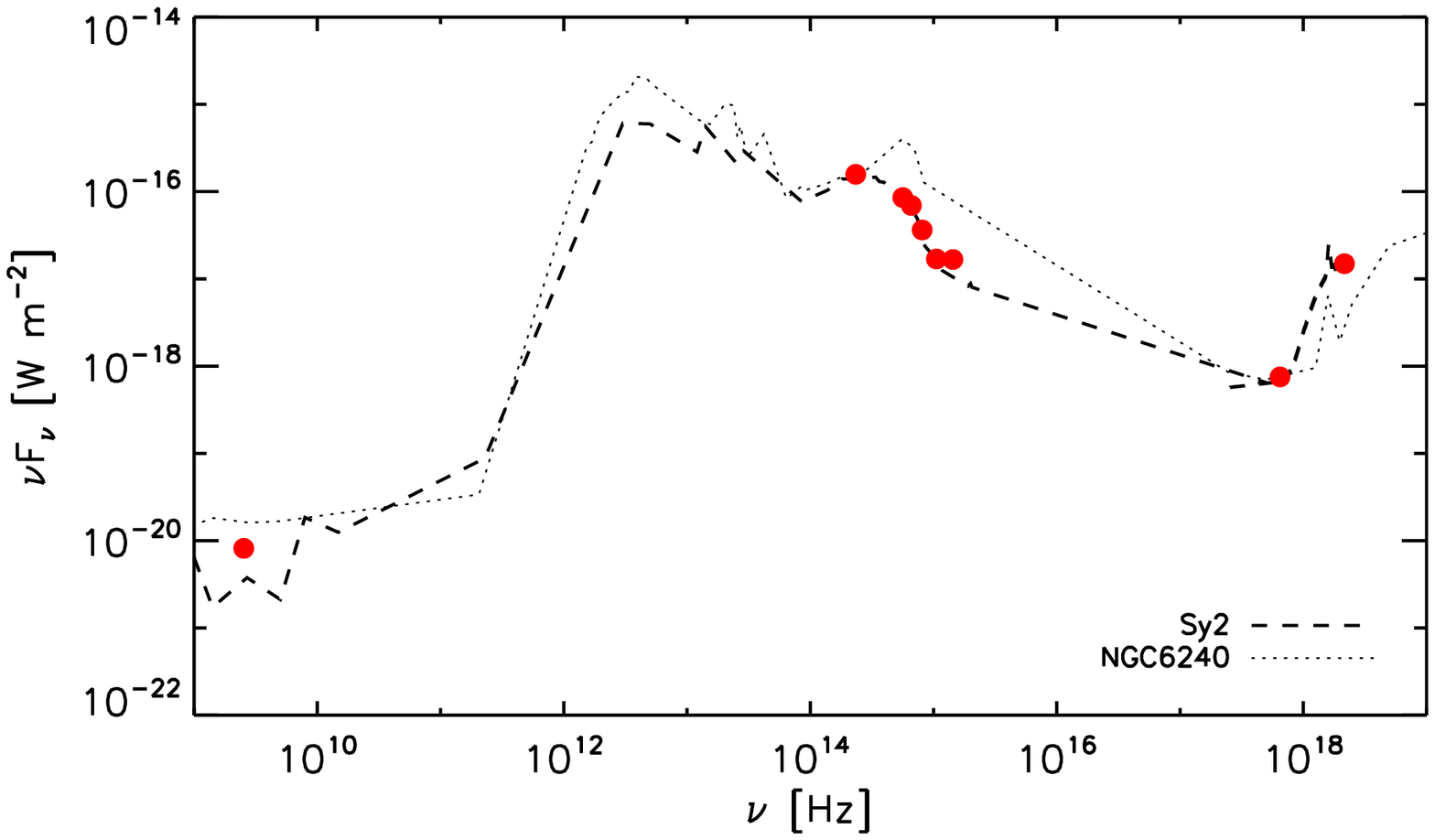}\\
\plottwo{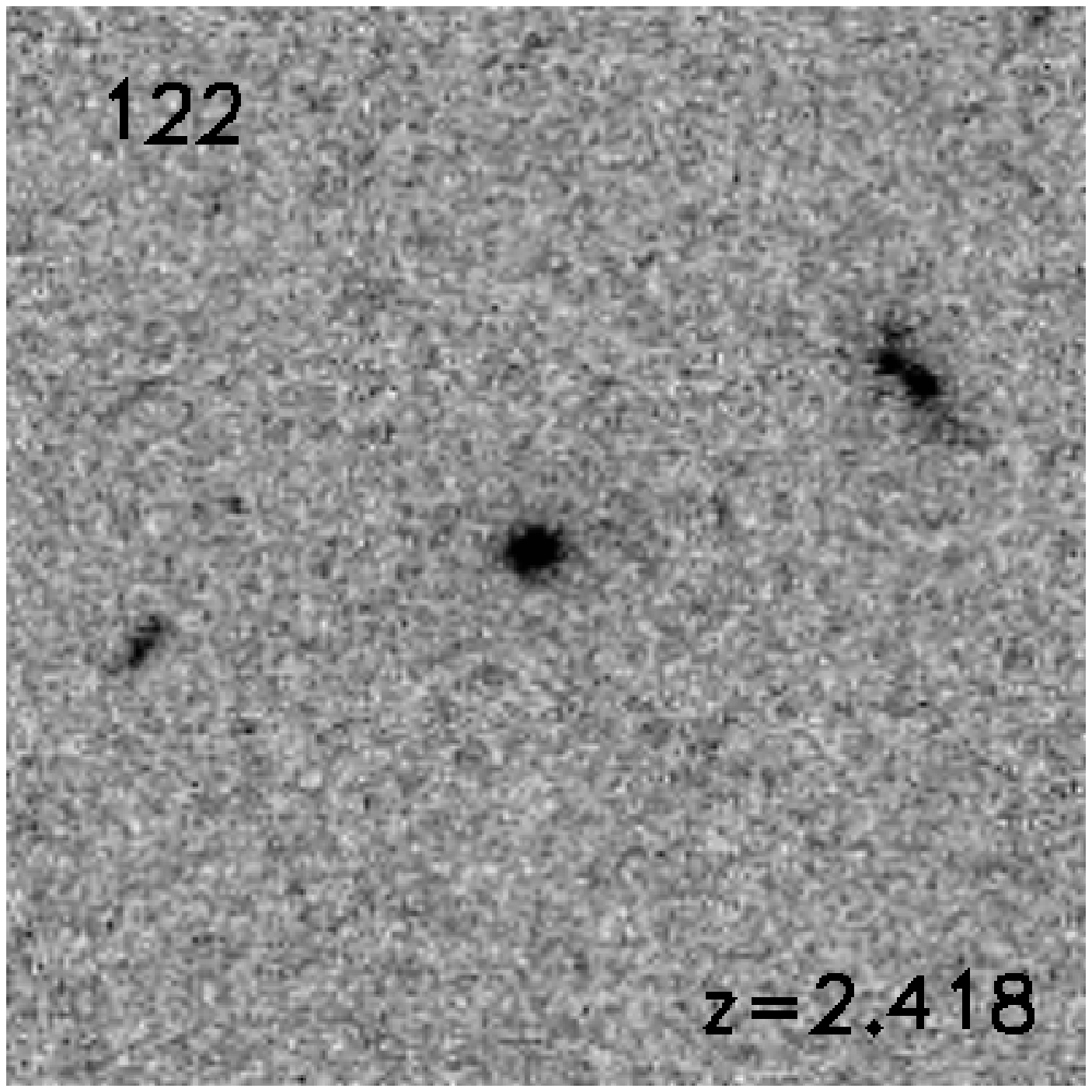}{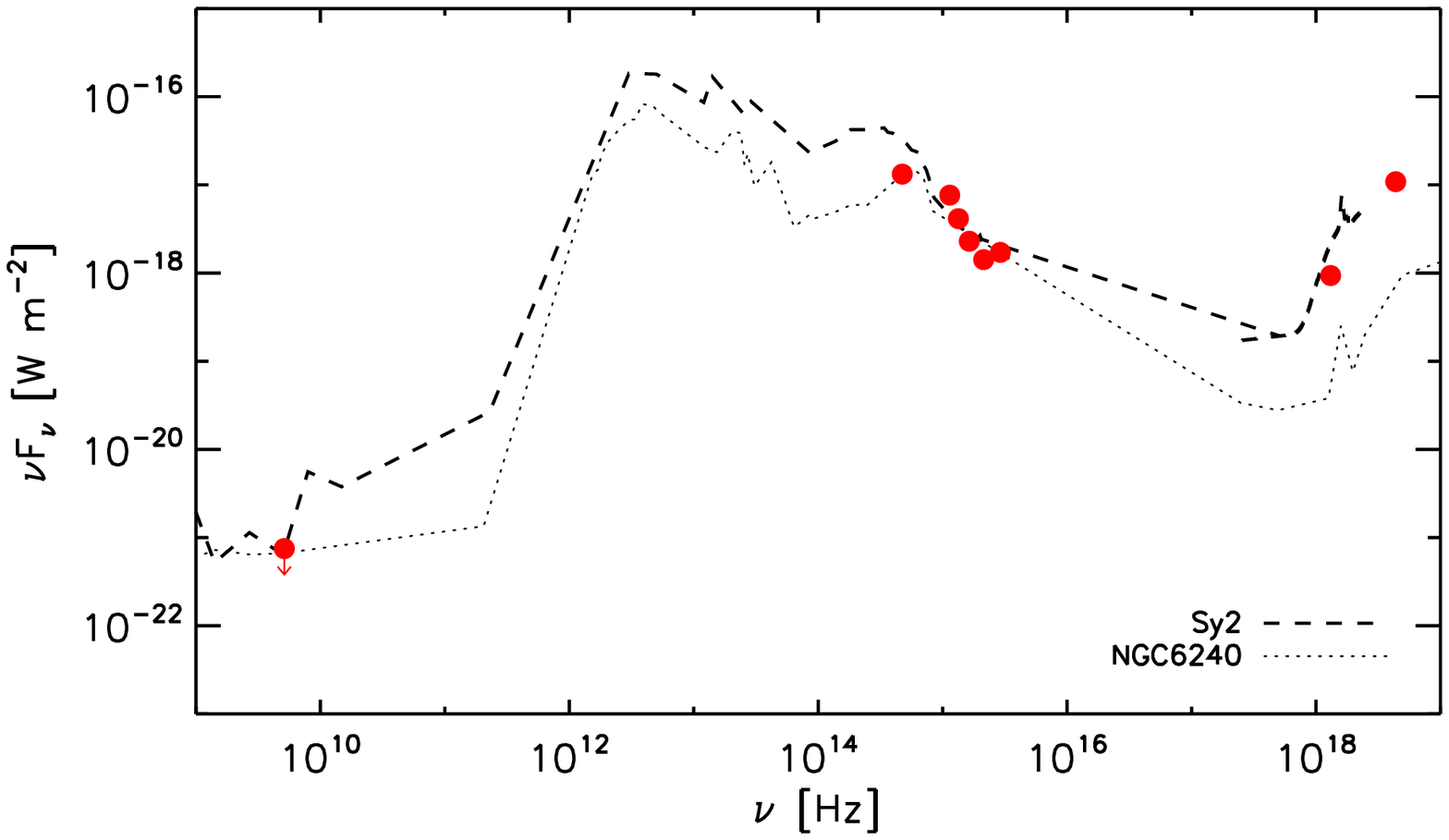}\\
\plottwo{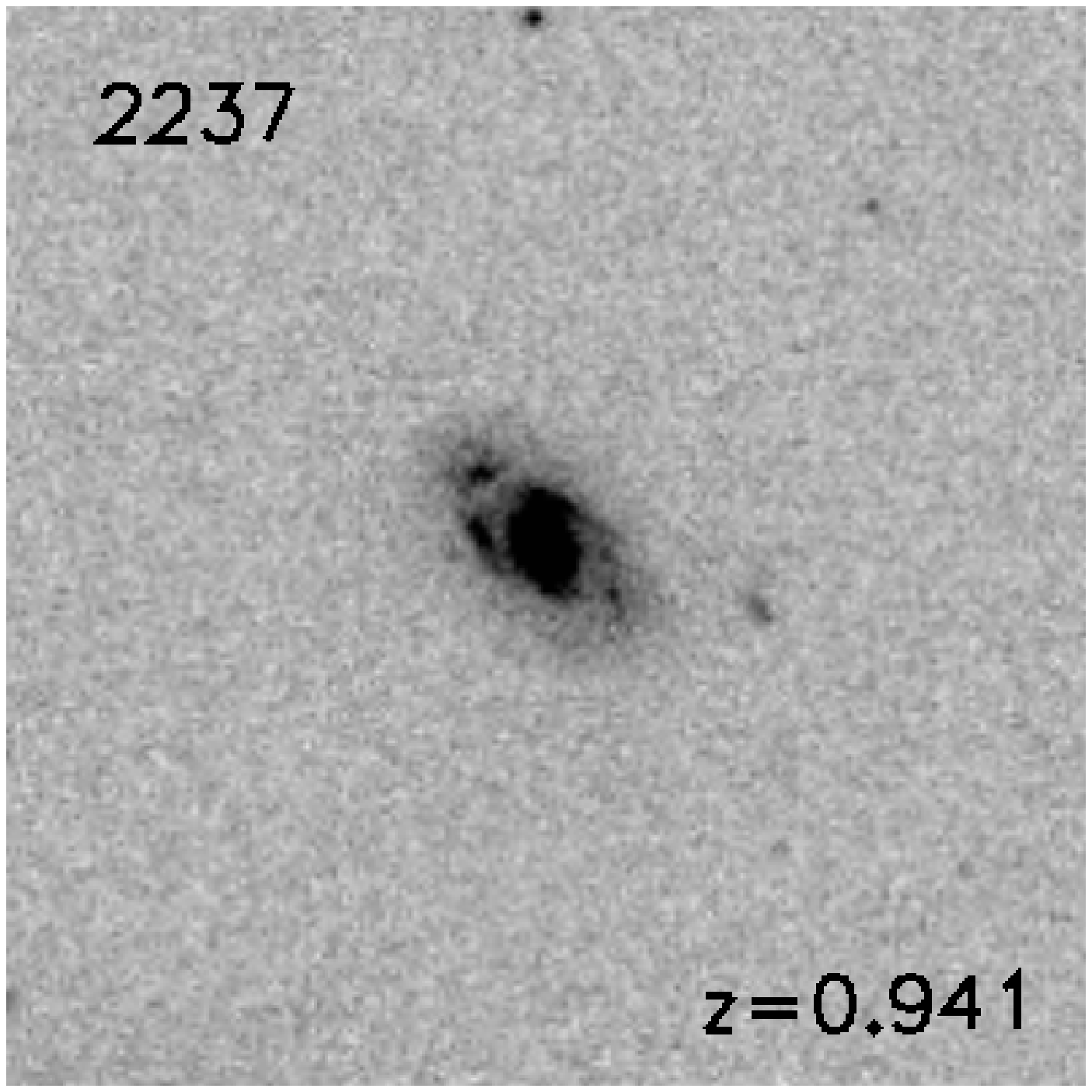}{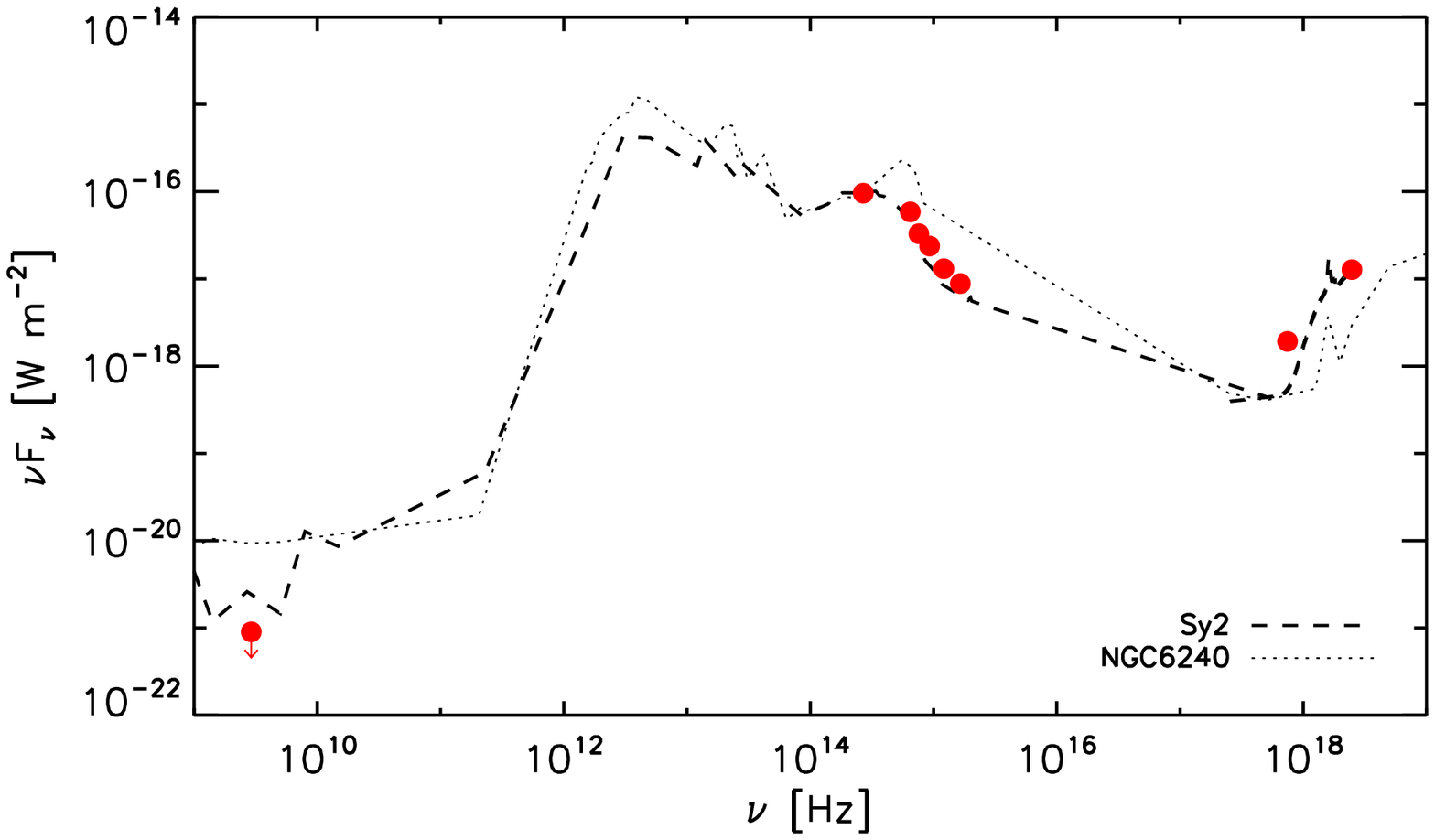}\\
\plottwo{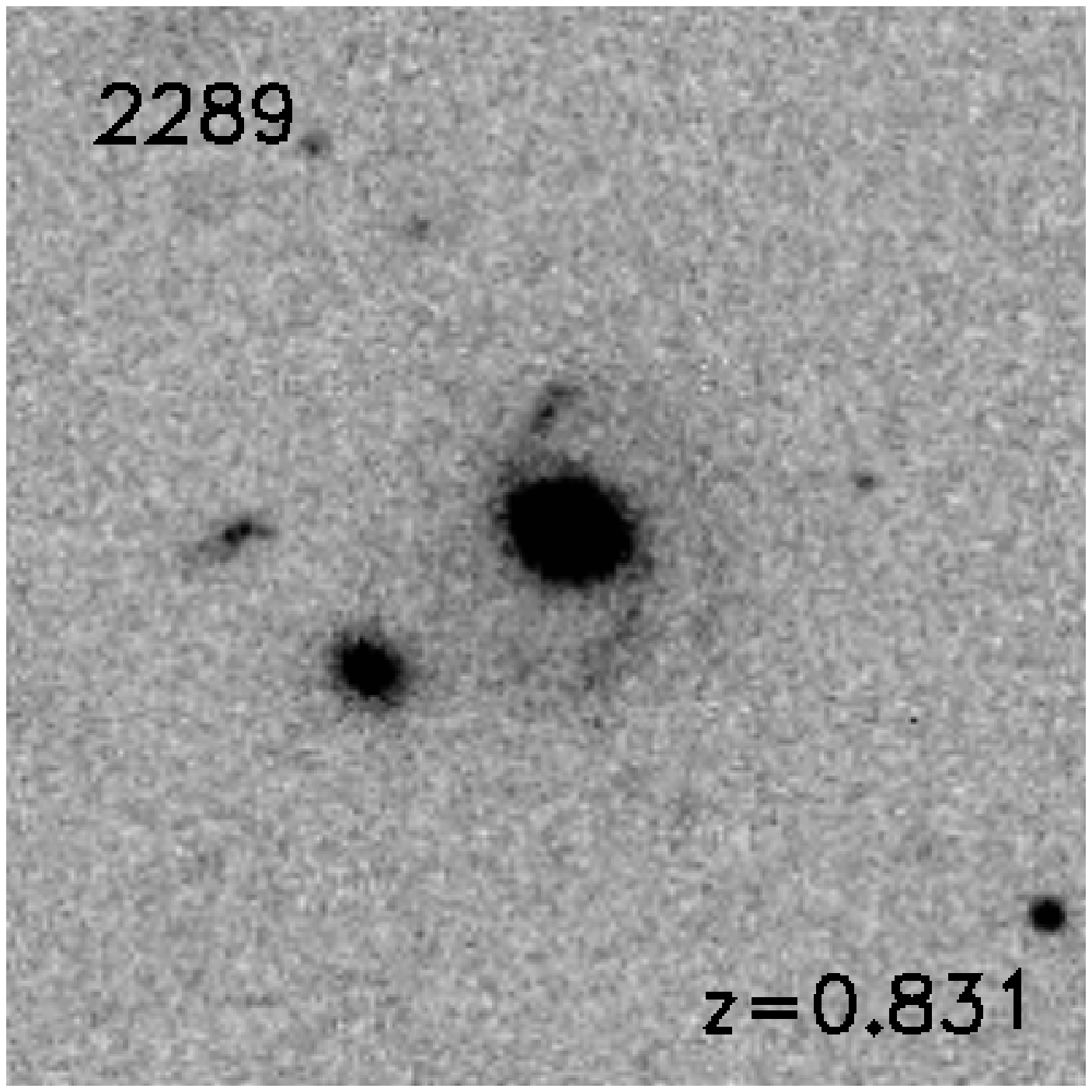}{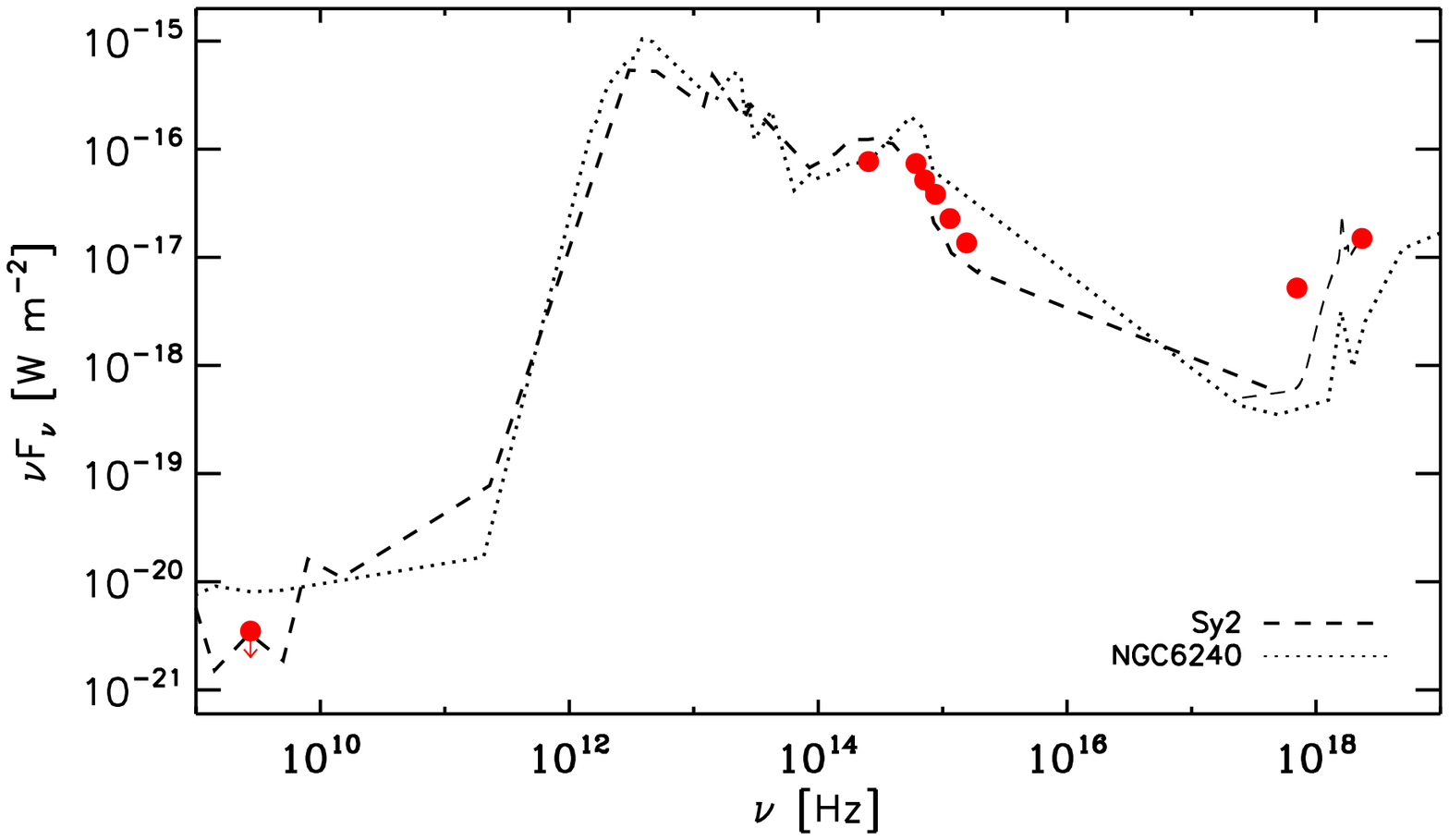}\\
\caption{{\it Left}: {\it i}-band (F775W) ACS cutouts for the four
  Type-2 QSO candidates. Each cutout is 10 arcsec across. {\it Right}:
  the spectral energy distribution of the Type-2 QSO candidates
  (filled circles) compared to the SED of a composite Seyfert-2
  spectrum (dashed line) and NGC6240 (dotted line).}
\label{qso2_sed}
\end{figure*}

Using the spectral parameters from the best-fit model, we correct
  the X-ray luminosity of each source for the intrinsic and Galactic
  absorption. These corrected luminosities are plotted in Fig.
\ref{lx_nh} versus the N$_{\rm H}$ for all the sources in our sample.
Four objects are characterized by a high X-ray luminosity (L$_{\rm X}
[0.5-10~keV] > 10^{44}$ erg s$^{-1}$) and substantial absorption
(N$_{\rm H} > 10^{22}$ cm$^{-2}$) and we can therefore classify them
as Type-2 QSOs. Radio-loud Type-2 QSOs are known since long times
thanks to radio surveys (see \citealt{mccarthy93} for a comprehensive
review), while radio-quiet Type-2 QSOs have been observed only
recently in {\it Chandra} and XMM-{\it Newton} X-ray surveys
\citep{dawson01,norman02,mainieri02,stern02,dellaceca03,fiore03,tozzi06}
and optical surveys (SDSS, \citealt{zakamska03}). Two of our Type-2
QSOs candidates, xid$=70, 2289$, are clearly detected in the radio at
20 cm using the Very Large Array (VLA) with an integrated flux of
$540\pm24$ and $52\pm11$ microJy respectively
\citep{schinnerer07}. The radio power of these two sources is therefore 
P$_{\rm 1.4 GHz}= 9.8 \times 10^{23}$ and $1.5 \times 10^{23}$
W/Hz. Historically such radio power has been used to divide radio loud
and radio quiet AGN but such a dividing line appears to be redshift
dependent: $\approx 5 \times 10^{23}$ W/Hz for the Palomar Green
sample (mainly below $z<0.3$) up to $5\times 10^{25}$ W/Hz for the
Large Bright Quasar Survey sample ($<z>\sim 1.2$). Since our two
sources are at z$\sim0.7-0.8$, we suggest to classify them as radio
quiet AGN. The other two objects (xid$= 122, 2237$) are not detected
in the radio and we can fix a 4.5 $\sigma$ upper limit to their radio
flux of 50 and 54 microJy
\citep{schinnerer07}.\\ The optical spectra of these four sources show
high excitation emission lines and their redshifts are 0.688, 0.831,
0.941, 2.418 respectively for xid 70, 2289, 2237, 122.\\ Using the
multi-band photometry available from the COSMOS survey
\citep{capak07}, we have derived the spectral energy distribution
(SED) for the four Type-2 QSOs and compared them with the spectrum of
NGC6240 and a Seyfert-2 composite spectrum derived from a sample of
local galaxies by \cite{schmitt97} and \cite{moran01}. While the SED
of NGC6240 does not reproduce well the observed photometry of our
Type-2 QSOs, an excellent description of the same is given by the
composite Seyfert-2 SED (see right panels in Fig. \ref{qso2_sed}).
Furthermore, the R-K colors of these four objects are red (R-K=4.58,
3.91, 4.97, 4.76 respectively) although they can not be classified as
EROs (R-K$>5$).



\section{Comparison between X-ray and optical classifications.}
\label{classification}

A classification based on the properties of the optical spectra of the
135 sources in our sample divides them into {\it 'Broad Line AGN'}
(BLAGN, 86 objects) if emission lines broader than 2000 km s$^{-1}$
are present, {\it 'Narrow Line AGN'} (NLAGN, 32 objects) if the
optical spectrum shows high excitation emission lines and {\it
  'galaxy'} (gal, 17 objects) if there is no sign of AGN activity from
the optical spectrum. As shown by deep {\it Chandra} and XMM-{\it
  Newton} surveys (e.g. \citealt{szokoly04}) a pure optical
classification of AGN is biased against absorbed sources that appear
as normal galaxies at those wavelengths. As previously done by
\cite{szokoly04} and \cite{tozzi06}, we introduce an X-ray based
classification: we define {\it X-ray absorbed} AGN sources that are
best fitted by an APL model compared to the PL one and have L$_{\rm
X}>10^{42}$ erg s$^{-1}$, {\it X-ray unabsorbed} AGN sources best
fitted with a PL model and L$_{\rm X}>10^{42}$ erg s$^{-1}$ and
finally {\it X-ray galaxies} sources with L$_{\rm X}<10^{42}$ erg
s$^{-1}$.\\ Table \ref{opt_xray} shows the comparison of the optical
and X-ray classifications for our 135 sources.  Ninety-one of these
sources ($\sim 67\%$) have a similar classification from the optical
and X-ray data. The best agreement between the two classifications is
for Broad Line AGN (optical) and X-ray unabsorbed AGN (X-ray) for
which the fractions of similar classifications are of the order of
$91\%$ ($78/86$ Broad Line AGN) and $76\%$ ($78/102$ X-ray unabsorbed
AGN) respectively. The $\sim 9\%$ of BL AGN that show X-ray absorption
in their X-ray spectra have values of the column density N$_{\rm H}$
below $10^{22}$ cm$^{-2}$ (see empty histogram in
Fig. \ref{nh_histo}). The main difference is instead for objects
classified as galaxies on the basis of the optical spectra.  Most of
these objects ($16/17$) are classified as AGN (11 absorbed and 5
unabsorbed) on the basis of the X-ray luminosity. This confirms that
the X-ray classification is more successful than the optical one in
revealing the presence of black hole activity. The situation is
intermediate for Narrow Line and X-ray absorbed AGN: only $\sim 41\%$
of the optically classified Narrow Line AGN do show detectable X-ray
absorption. We note that of the remaining Narrow Line AGN, $\sim 80\%$
have z$>0.4$ and therefore the H$\alpha$ line is outside the observed
wavelength range, while for nine of them the MgII line is inside the
observed range (i.e. $0.92<z<2.29$) but the S/N of the spectra could
not be sufficient to detect a weak broad line. It is therefore
possible that at least part of the disagreement between the optical
and the X-ray classifications for these objects is due to less than
optimal optical spectra, in terms of either spectral coverage or S/N.


\section{Conclusions}

We have presented the detailed spectral analysis of 135 X-ray sources
from the XMM-{\it Newton} wide-field survey in the COSMOS field. All
the sources in our sample have more than 100 net counts in the
[0.3-10] keV energy band and have been spectroscopically
identified. For each source we have performed an accurate spectral fit
in order to measure the continuum shape, the amount of absorbing
matter and the strength of other spectral features. Our main results
are summarized as follows:

\begin{itemize}

\item We find that, to the X-ray flux limit we are sampling (F$_{\rm
    X}[0.5-10]=1.4 \times 10^{-15}$ erg cm$^{-2}$ s$^{-1}$), $\sim
  76\%$ of the spectra are well reproduced with a single power-law
  model, $\sim 20\%$ require an absorbed power-law model and the
  remaining $\sim 4\%$ need more complex models.

\item The average value of the spectral slope of the intrinsic spectrum for
  the 82 sources with more than 180 net counts ({\it Sample-1}) is
  $<\Gamma> = 2.06\pm0.08$ with an intrinsic dispersion of
  $\sigma_{int}=0.24$.

\item We find no correlation between the spectral slope $\Gamma$ and
  the amount of intrinsic absorption N$_{\rm H}$, confirming that the
  hardening of the X-ray spectra going to fainter X-ray fluxes is due
  to the increased fraction of absorbed X-ray sources.

\item None of the X-ray sources with a column density N$_{\rm H}>10^{22}$
  cm$^{-2}$ shows broad line in their optical spectra, although a
  fraction ($9\%$) of broad line AGN shows intrinsic absorption in
  excess to the Galactic value.

\item We detect (at more than $90\%$ confidence level) the Fe
  K$\alpha$ line in three objects. One of them is well described by a
  pure reflection model plus a Gaussian line at 6.4 keV rest-frame.
  This, the large equivalent width of the Fe line (although with large
  uncertainties) and diagnostics based on lines ratios from the optical
  spectrum support the hypothesis that this particular source is a
  Compton thick AGN.

\item We find four radio-quiet Type-2 QSOs. Their spectral energy distribution is well reproduced with a Seyfert-2
composite spectrum.


\item We confirm that in order to have a less biased sample of AGN it
  is crucial to complement the optical spectral properties with the
  X-ray informations (L$_{\rm X}$ and N$_{\rm H}$), since many
  apparently normal galaxies in the optical band are instead absorbed
  AGN.

\end{itemize}

This is the first work on the X-ray spectral properties of the AGN in
the COSMOS survey. We remark that once the XMM-{\it Newton}
observations will be completed and the planned spectroscopic follow-up
finished, we will be able to analyze the X-ray spectral properties of
the AGN on a much larger sample and compare them with the properties
of the AGN/host-galaxies at almost all the wavelengths.

\acknowledgments 

This work is based on observations obtained with XMM--{\it Newton}, an
ESA science mission with instruments and contributions directly funded
by ESA Member States and the US (NASA).  In Germany, the XMM-Newton
project is supported by the Bundesministerium für Wirtschaft und
Technologie/Deutsches Zentrum für Luft- und Raumfahrt (BMWI/DLR, FKZ
50 OX 0001), the Max-Planck Society and the Heidenhain-Stiftung. Part
of this work was supported by the Deutsches Zentrum f\"ur Luft-- und
Raumfahrt, DLR project numbers 50 OR 0207 and 50 OR 0405. In Italy,
the XMM-COSMOS project is supported by INAF and MIUR under grants
PRIN/270/2003 and Cofin-03-02-23. We acknowledge financial
contribution from contract ASI-INAF I/023/05/0.  Based on observations
with the NASA/ESA {\em Hubble Space Telescope}, obtained at the Space
Telescope Science Institute, which is operated by AURA Inc, under NASA
contract NAS 5-26555. The HST COSMOS Treasury program was supported
through NASA grant HST-GO-09822. Also based on data collected at : the
Subaru Telescope, which is operated by the National Astronomical
Observatory of Japan; the European Southern Observatory under Large
Program 175.A-0839, Chile; the National Radio Astronomy Observatory
which is a facility of the National Science Foundation operated under
cooperative agreement by Associated Universities, Inc.\\ We are
grateful to Paolo Tozzi, Kazushi Iwasawa and Paolo Padovani for
inspiring discussions.  We gratefully acknowledge the entire COSMOS
collaboration consisting of more than 70 scientists.  More information
on the COSMOS survey is available at\\
\url{http://www.astro.caltech.edu/$\sim$cosmos}. It is a pleasure the
acknowledge the excellent services provided by the NASA IPAC/IRSA
staff (Anastasia Laity, Anastasia Alexov, Bruce Berriman and John
Good) in providing online archive and server capabilities for the
COSMOS datasets.

 \clearpage
 
 


\clearpage
\LongTables
\begin{landscape}
\begin{deluxetable}{crrrrrcccrrrlll}
\tabletypesize{\tiny}
\tablecaption{X-ray spectral fit parameters \label{xspec_catalogue}}
\tablewidth{0pt}
\tablehead{
IAU\tablenotemark{a} & XID\tablenotemark{b}  &  \multicolumn{1}{c}{RA\tablenotemark{c}} &  \multicolumn{1}{c}{Dec\tablenotemark{c}} &  \multicolumn{1}{c}{counts\tablenotemark{d}} &  \multicolumn{1}{c}{z\tablenotemark{e}} &  \multicolumn{1}{c}{MODEL\tablenotemark{f}} &  \multicolumn{1}{c}{$\Gamma$} &  \multicolumn{1}{c}{N$_{\rm H}$} &  \multicolumn{1}{c}{fx\tablenotemark{g}} &  \multicolumn{1}{c}{fx\tablenotemark{g}} &  \multicolumn{1}{c}{fx\tablenotemark{g}} & \multicolumn{1}{c}{L$_{\rm X}$\tablenotemark{h}} & \multicolumn{1}{c}{L$_{\rm X}$\tablenotemark{h}} & \multicolumn{1}{c}{L$_{\rm X}$\tablenotemark{h}} \\
 & & \multicolumn{2}{c}{(J2000)} & [0.3-10] & & & & & [0.5-2] & [2-10] & [0.5-10]  & [0.5-2] & [2-10] & [0.5-10]\\
}
\startdata
XMMC$\_$J100025.30+015851.2 & 1 & 10:00:25.30 & 1:58:51.19 & 4396 & 0.373 & PL & 2.11$^{2.15}_{2.08}$ & ..... & 746.88 & 795.94 & 1542.80 & 43.55 & 43.58 & 43.87 \\
XMMC$\_$J095857.50+021314.1 & 2 & 9:58:57.50 & 2:13:14.10 & 1896 & 1.024 & PL & 1.79$^{1.84}_{1.75}$ & ..... & 1178.70 & 2004.80 & 3183.60 & 44.81 & 45.05 & 45.25 \\
XMMC$\_$J095902.83+021906.8 & 3 & 9:59:02.83 & 2:19:06.77 & 2683 & 0.345 & PL & 2.07$^{2.12}_{2.03}$ & ..... & 1318.50 & 1484.60 & 2803.10 & 43.72 & 43.77 & 44.05 \\
XMMC$\_$J095858.68+021458.1 & 4 & 9:58:58.68 & 2:14:58.12 & 1188 & 0.132 & PL & 1.78$^{1.84}_{1.71}$ & ..... & 851.43 & 1491.50 & 2342.90 & 42.59 & 42.84 & 43.03 \\
XMMC$\_$J095918.91+020951.3 & 5 & 9:59:18.91 & 2:09:51.26 & 1517 & 1.154 & PL & 1.78$^{1.84}_{1.72}$ & ..... & 377.01 & 659.48 & 1036.50 & 44.45 & 44.69 & 44.89 \\
XMMC$\_$J100043.26+020636.6 & 6 & 10:00:43.26 & 2:06:36.56 & 1664 & 0.360 & PL & 2.18$^{2.26}_{2.11}$ & ..... & 388.08 & 377.01 & 765.09 & 43.25 & 43.23 & 43.54 \\
XMMC$\_$J100013.02+023521.8 & 8 & 10:00:13.02 & 2:35:21.82 & 1121 & 0.699 & PL & 2.45$^{2.53}_{2.38}$ & ..... & 627.74 & 407.25 & 1035.00 & 44.13 & 43.95 & 44.35 \\
XMMC$\_$J095940.86+021938.6 & 9 & 9:59:40.86 & 2:19:38.56 & 1094 & 1.459 & PL & 1.99$^{2.08}_{1.91}$ & ..... & 196.07 & 249.26 & 445.33 & 44.42 & 44.52 & 44.77 \\
XMMC$\_$J100034.95+020234.0 & 11 & 10:00:34.95 & 2:02:34.03 & 789 & 1.177 & PL & 2.25$^{2.37}_{2.13}$ & ..... & 139.90 & 122.46 & 262.35 & 44.04 & 43.98 & 44.31 \\
XMMC$\_$J100049.95+020500.0 & 12 & 10:00:49.95 & 2:05:00.03 & 741 & 1.235 & PL & 2.50$^{2.63}_{2.38}$ & ..... & 240.22 & 144.84 & 385.06 & 44.33 & 44.11 & 44.53 \\
XMMC$\_$J095924.69+015954.5 & 17 & 9:59:24.69 & 1:59:54.45 & 1771 & 1.236 & PL & 2.23$^{2.30}_{2.16}$ & ..... & 728.21 & 653.76 & 1382.00 & 44.81 & 44.76 & 45.09 \\
XMMC$\_$J095958.60+021531.0 & 19 & 9:59:58.60 & 2:15:31.02 & 487 & 0.658 & PL & 2.05$^{2.20}_{1.90}$ & ..... & 247.59 & 278.06 & 525.65 & 43.63 & 43.70 & 43.97 \\
XMMC$\_$J100058.80+022556.7 & 20 & 10:00:58.80 & 2:25:56.68 & 575 & 0.693 & PL & 2.22$^{2.34}_{2.10}$ & ..... & 182.66 & 165.75 & 348.41 & 43.62 & 43.58 & 43.90 \\
XMMC$\_$J100055.46+023442.0 & 21 & 10:00:55.46 & 2:34:41.99 & 571 & 1.403 & PL & 2.15$^{2.28}_{2.02}$ & ..... & 158.63 & 159.81 & 318.44 & 44.28 & 44.29 & 44.58 \\
XMMC$\_$J100046.85+020405.2 & 22 & 10:00:46.85 & 2:04:05.25 & 586 & 0.552 & PL & 2.70$^{2.84}_{2.56}$ & ..... & 113.30 & 51.61 & 164.91 & 43.11 & 42.77 & 43.27 \\
XMMC$\_$J095909.63+021917.2 & 23 & 9:59:09.63 & 2:19:17.22 & 891 & 0.378 & PL & 2.05$^{2.16}_{1.95}$ & ..... & 314.47 & 360.31 & 674.78 & 43.17 & 43.23 & 43.50 \\
XMMC$\_$J100024.74+023148.3 & 24 & 10:00:24.74 & 2:31:48.34 & 382 & 1.318 & PL & 2.65$^{2.79}_{2.51}$ & ..... & 293.47 & 143.55 & 437.02 & 44.48 & 44.17 & 44.66 \\
XMMC$\_$J100024.55+020618.5 & 25 & 10:00:24.55 & 2:06:18.48 & 440 & 2.281 & PL & 1.75$^{1.88}_{1.62}$ & ..... & 156.54 & 286.15 & 442.68 & 44.79 & 45.06 & 45.25 \\
XMMC$\_$J095949.51+020139.1 & 30 & 9:59:49.51 & 2:01:39.09 & 611 & 1.758 & PL & 2.51$^{2.67}_{2.35}$ & ..... & 120.90 & 72.33 & 193.24 & 44.41 & 44.18 & 44.61 \\
XMMC$\_$J095947.05+022209.4 & 31 & 9:59:47.05 & 2:22:09.38 & 700 & 0.909 & PL & 2.27$^{2.40}_{2.15}$ & ..... & 128.42 & 108.14 & 236.56 & 43.72 & 43.65 & 43.99 \\
XMMC$\_$J100114.36+022357.5 & 33 & 10:01:14.36 & 2:23:57.47 & 410 & 1.799 & PL & 2.34$^{2.50}_{2.19}$ & ..... & 119.16 & 91.12 & 210.28 & 44.42 & 44.31 & 44.67 \\
XMMC$\_$J095958.62+021805.9 & 34 & 9:59:58.62 & 2:18:05.92 & 521 & 1.792 & PL & 1.99$^{2.14}_{1.85}$ & ..... & 111.88 & 142.74 & 254.62 & 44.39 & 44.50 & 44.75 \\
XMMC$\_$J095928.45+022107.6 & 35 & 9:59:28.45 & 2:21:07.64 & 440 & 0.346 & PL & 2.54$^{2.68}_{2.40}$ & ..... & 117.30 & 67.04 & 184.34 & 42.72 & 42.47 & 42.91 \\
XMMC$\_$J095940.18+022306.3 & 37 & 9:59:40.18 & 2:23:06.28 & 698 & 1.132 & PL & 2.16$^{2.29}_{2.06}$ & ..... & 89.97 & 89.38 & 179.35 & 43.81 & 43.80 & 44.10 \\
XMMC$\_$J100058.94+015359.5 & 38 & 10:00:58.94 & 1:53:59.45 & 369 & 1.559 & APL & 2.04$^{2.19}_{1.85}$ & 20.73$^{21.46}_{20.42}$ & 106.94 & 129.53 & 236.47 & 44.32 & 44.39 & 44.65 \\
XMMC$\_$J100114.94+020208.9 & 40 & 10:01:14.94 & 2:02:08.93 & 602 & 0.989 & PL & 2.01$^{2.14}_{1.89}$ & ..... & 465.40 & 574.11 & 1039.50 & 44.37 & 44.47 & 44.72 \\
XMMC$\_$J100025.43+020734.4 & 41 & 10:00:25.43 & 2:07:34.43 & 315 & 0.114 & APL+po & 1.95$^{2.37}_{1.65}$ & 21.51$^{21.55}_{21.27}$ & 76.89 & 204.98 & 281.87 & 41.63 & 41.76 & 42.00 \\
XMMC$\_$J100202.80+022435.8 & 42 & 10:02:02.80 & 2:24:35.82 & 476 & 0.988 & PL & 2.15$^{2.32}_{1.99}$ & ..... & 197.35 & 197.84 & 395.20 & 44.00 & 44.00 & 44.30 \\
XMMC$\_$J100051.57+021215.8 & 44 & 10:00:51.57 & 2:12:15.80 & 305 & 1.829 & PL & 2.14$^{2.32}_{1.99}$ & ..... & 105.65 & 107.47 & 213.13 & 44.39 & 44.40 & 44.69 \\
XMMC$\_$J100014.12+020054.2 & 51 & 10:00:14.12 & 2:00:54.18 & 336 & 2.497 & PL & 1.98$^{2.26}_{1.88}$ & ..... & 50.78 & 65.43 & 116.21 & 44.44 & 44.55 & 44.80 \\
XMMC$\_$J100016.35+015104.3 & 52 & 10:00:16.35 & 1:51:04.30 & 297 & 1.135 & PL & 1.85$^{2.00}_{1.70}$ & ..... & 107.96 & 169.84 & 277.80 & 43.82 & 44.02 & 44.23 \\
XMMC$\_$J100131.15+022924.8 & 54\tablenotemark{i} & 10:01:31.15 & 2:29:24.82 & 246 & 0.350 & R-S & ..... & ..... & 90.18 & 24.44 & 114.62 & 42.57 & 42.00 & 42.67 \\\vspace{1 mm}
XMMC$\_$J100001.16+021413.9 & 56 & 10:00:01.16 & 2:14:13.92 & 110 & 1.407 & PL & 2.00 & ..... & 89.25 & 101.56 & 190.81 & 44.04 & 44.09 & 44.37 \\
XMMC$\_$J100047.09+020017.7 & 59 & 10:00:47.09 & 2:00:17.71 & 226 & 1.904 & PL & 2.12$^{2.33}_{1.93}$ & ..... & 59.74 & 63.30 & 123.04 & 44.18 & 44.21 & 44.50 \\
XMMC$\_$J095907.84+020819.3 & 63 & 9:59:07.84 & 2:08:19.34 & 264 & 0.354 & PL & 1.95$^{2.19}_{1.73}$ & ..... & 193.50 & 260.69 & 454.19 & 42.85 & 42.98 & 43.22 \\
XMMC$\_$J095934.63+020627.9 & 64 & 9:59:34.63 & 2:06:27.94 & 299 & 0.686 & PL & 1.64$^{1.81}_{1.47}$ & ..... & 56.23 & 121.35 & 177.58 & 43.00 & 43.34 & 43.50 \\
XMMC$\_$J100041.87+022411.1 & 65 & 10:00:41.87 & 2:24:11.07 & 122 & 0.979 & PL & 2.00 & ..... & 61.79 & 77.64 & 139.44 & 43.49 & 43.59 & 43.84 \\
XMMC$\_$J095928.45+021950.5 & 66 & 9:59:28.45 & 2:19:50.47 & 436 & 1.488 & PL & 2.22$^{2.40}_{2.04}$ & ..... & 81.58 & 74.39 & 155.96 & 44.06 & 44.02 & 44.34 \\
XMMC$\_$J100137.74+022845.1 & 67 & 10:01:37.74 & 2:28:45.09 & 224 & 0.367 & PL & 1.93$^{2.16}_{1.71}$ & ..... & 69.35 & 97.21 & 166.56 & 42.37 & 42.51 & 42.75 \\
XMMC$\_$J095934.92+021028.5 & 69 & 9:59:34.92 & 2:10:28.46 & 133 & 2.412 & PL & 2.00 & ..... & 78.25 & 98.32 & 176.57 & 44.55 & 44.65 & 44.91 \\
XMMC$\_$J100036.13+022830.7 & 70 & 10:00:36.13 & 2:28:30.66 & 181 & 0.688 & APL & 2.78$^{3.92}_{2.27}$ & 23.23$^{23.41}_{23.04}$ & 25.19 & 579.88 & 605.07 & 44.68 & 44.29 & 44.83 \\
XMMC$\_$J100129.81+023239.6 & 72 & 10:01:29.81 & 2:32:39.56 & 220 & 0.825 & APL & 1.72$^{2.04}_{1.40}$ & 21.00$^{21.47}_{20.42}$ & 55.72 & 114.19 & 169.91 & 43.33 & 43.61 & 43.80 \\
XMMC$\_$J100031.66+014757.4 & 75 & 10:00:31.66 & 1:47:57.40 & 363 & 1.681 & PL & 1.94$^{2.10}_{1.80}$ & ..... & 154.44 & 211.03 & 365.48 & 44.46 & 44.60 & 44.84 \\
XMMC$\_$J100028.71+021744.5 & 78 & 10:00:28.71 & 2:17:44.48 & 203 & 1.039 & PL & 1.72$^{2.02}_{1.45}$ & ..... & 50.69 & 95.86 & 146.54 & 43.46 & 43.74 & 43.93 \\
XMMC$\_$J100124.93+022032.2 & 79 & 10:01:24.93 & 2:20:32.19 & 171 & 1.708 & PL & 2.00 & ..... & 74.53 & 93.65 & 168.18 & 44.16 & 44.26 & 44.52 \\
XMMC$\_$J100105.65+015603.0 & 81 & 10:01:05.65 & 1:56:03.04 & 285 & 0.915 & APL & 1.44$^{1.65}_{1.23}$ & 21.44$^{21.67}_{20.94}$ & 50.50 & 159.64 & 210.14 & 43.26 & 43.70 & 43.84 \\
XMMC$\_$J100117.73+023309.0 & 85 & 10:01:17.73 & 2:33:09.02 & 184 & 1.001 & APL & 1.99$^{2.42}_{1.58}$ & 21.16$^{21.63}_{20.42}$ & 50.33 & 68.88 & 119.21 & 43.50 & 43.60 & 43.85 \\
XMMC$\_$J100048.01+021128.0 & 94 & 10:00:48.01 & 2:11:28.00 & 142 & 1.515 & PL & 2.00 & ..... & 80.93 & 101.69 & 182.61 & 44.07 & 44.17 & 44.43 \\
XMMC$\_$J100136.47+025304.5 & 96 & 10:01:36.47 & 2:53:04.50 & 134 & 2.117 & PL & 2.00 & ..... & 163.26 & 201.31 & 364.57 & 44.73 & 44.83 & 45.08 \\
XMMC$\_$J100031.41+022819.2 & 101 & 10:00:31.41 & 2:28:19.18 & 131 & 0.926 & PL & 2.00 & ..... & 47.26 & 59.38 & 106.63 & 43.31 & 43.41 & 43.66 \\
XMMC$\_$J100028.20+015547.0 & 103 & 10:00:28.20 & 1:55:46.98 & 144 & 1.519 & PL & 2.00 & ..... & 53.96 & 67.80 & 121.76 & 43.90 & 44.00 & 44.25 \\
XMMC$\_$J100038.13+022455.8 & 106 & 10:00:38.13 & 2:24:55.79 & 141 & 0.710 & APL+po & 2.00 & 22.33$^{22.66}_{21.98}$ & 14.54 & 43.92 & 58.47 & 42.85 & 42.95 & 43.20 \\
XMMC$\_$J095935.73+020537.2 & 113 & 9:59:35.73 & 2:05:37.24 & 101 & 1.910 & PL & 2.00 & ..... & 55.61 & 69.87 & 125.48 & 44.16 & 44.26 & 44.51 \\
XMMC$\_$J100210.73+023028.0 & 115 & 10:02:10.73 & 2:30:27.97 & 591 & 1.161 & APL & 2.10$^{2.23}_{1.95}$ & 21.10$^{21.42}_{20.42}$ & 176.22 & 204.96 & 381.18 & 44.18 & 44.22 & 44.50 \\
XMMC$\_$J100049.61+021709.2 & 116 & 10:00:49.61 & 2:17:09.17 & 218 & 0.874 & APL & 2.11$^{2.38}_{1.61}$ & 21.11$^{21.58}_{20.42}$ & 74.38 & 87.32 & 161.71 & 43.53 & 43.56 & 43.84 \\
XMMC$\_$J100013.45+021400.5 & 117 & 10:00:13.45 & 2:14:00.47 & 111 & 0.936 & APL$+$po & 2.00 & 22.76$^{23.14}_{22.37}$ & 28.01 & 35.20 & 63.21 & 43.09 & 43.19 & 43.45 \\
XMMC$\_$J100122.23+021334.0 & 119 & 10:01:22.23 & 2:13:33.99 & 328 & 0.891 & APL & 1.69$^{2.04}_{1.40}$ & 21.01$^{21.48}_{20.42}$ & 213.98 & 452.80 & 666.78 & 43.97 & 44.27 & 44.44 \\
XMMC$\_$J095945.47+021029.9 & 122 & 9:59:45.47 & 2:10:29.88 & 130 & 2.418 & APL & 2.00 & 23.69$^{23.91}_{23.44}$ & 14.90 & 147.41 & 162.32 & 44.83 & 44.93 & 45.18 \\
XMMC$\_$J100131.93+023335.5 & 123 & 10:01:31.93 & 2:33:35.46 & 142 & 2.065 & PL & 2.00 & ..... & 23.93 & 30.07 & 54.00 & 43.87 & 43.97 & 44.23 \\
XMMC$\_$J100001.27+022320.7 & 127 & 10:00:01.27 & 2:23:20.69 & 217 & 1.846 & PL & 2.55$^{2.91}_{2.22}$ & ..... & 34.62 & 19.63 & 54.25 & 43.91 & 43.67 & 44.11 \\
XMMC$\_$J100047.85+020756.1 & 128 & 10:00:47.85 & 2:07:56.15 & 120 & 2.161 & PL & 2.00 & ..... & 32.52 & 40.86 & 73.38 & 44.06 & 44.15 & 44.41 \\
XMMC$\_$J100100.90+015946.7 & 129 & 10:01:00.90 & 1:59:46.69 & 184 & 1.170 & PL & 1.98$^{2.21}_{1.76}$ & ..... & 196.94 & 256.78 & 453.72 & 44.13 & 44.25 & 44.49 \\
XMMC$\_$J100105.36+021348.0 & 133 & 10:01:05.36 & 2:13:47.96 & 144 & 2.627 & PL & 2.00 & ..... & 56.73 & 71.28 & 128.00 & 44.50 & 44.60 & 44.86 \\
XMMC$\_$J100011.78+021919.9 & 134 & 10:00:11.78 & 2:19:19.86 & 141 & 0.625 & PL & 2.00 & ..... & 35.13 & 44.14 & 79.28 & 42.76 & 42.86 & 43.12 \\
XMMC$\_$J095949.98+020010.6 & 137 & 9:59:49.98 & 2:00:10.57 & 195 & 1.808 & PL & 1.87$^{2.31}_{1.48}$ & ..... & 44.24 & 92.96 & 137.20 & 44.06 & 44.25 & 44.47 \\
XMMC$\_$J100033.55+015236.3 & 141 & 10:00:33.55 & 1:52:36.34 & 102 & 0.831 & APL & 2.00 & 21.64$^{21.96}_{21.19}$ & 51.21 & 85.40 & 136.61 & 43.36 & 43.45 & 43.71 \\
XMMC$\_$J100013.46+022656.7 & 143 & 10:00:13.46 & 2:26:56.66 & 140 & 0.732 & APL & 2.00 & 22.68$^{23.06}_{22.38}$ & 29.41 & 194.21 & 223.62 & 43.62 & 43.72 & 43.97 \\
XMMC$\_$J095938.49+020447.5 & 146 & 9:59:38.49 & 2:04:47.51 & 167 & 2.804 & APL & 2.00 & 21.95$^{22.26}_{21.23}$ & 53.36 & 73.96 & 127.32 & 44.59 & 44.69 & 44.94 \\
XMMC$\_$J100053.93+021614.2 & 147 & 10:00:53.93 & 2:16:14.22 & 112 & 2.944 & PL & 2.00 & ..... & 26.82 & 33.70 & 60.52 & 44.30 & 44.40 & 44.65 \\
XMMC$\_$J100052.57+021643.8 & 148 & 10:00:52.57 & 2:16:43.80 & 111 & 0.843 & PL & 2.00 & ..... & 41.34 & 51.94 & 93.27 & 43.15 & 43.25 & 43.50 \\
XMMC$\_$J100124.00+021446.4 & 152 & 10:01:24.00 & 2:14:46.45 & 172 & 0.894 & PL & 2.00 & ..... & 88.13 & 110.74 & 198.87 & 43.54 & 43.64 & 43.90 \\
XMMC$\_$J100108.44+022342.6 & 153 & 10:01:08.44 & 2:23:42.58 & 142 & 1.928 & APL & 2.00 & 21.91$^{22.20}_{21.40}$ & 33.00 & 49.24 & 82.24 & 44.02 & 44.12 & 44.37 \\
XMMC$\_$J100108.59+020053.2 & 161 & 10:01:08.59 & 2:00:53.24 & 254 & 2.681 & PL & 1.69$^{1.93}_{1.46}$ & ..... & 75.27 & 149.97 & 225.25 & 44.65 & 44.95 & 45.12 \\
XMMC$\_$J100118.55+015543.6 & 164 & 10:01:18.55 & 1:55:43.59 & 291 & 0.528 & PL & 2.54$^{2.73}_{2.36}$ & ..... & 176.15 & 100.11 & 276.25 & 43.32 & 43.07 & 43.51 \\
XMMC$\_$J100043.30+021352.7 & 165 & 10:00:43.30 & 2:13:52.65 & 120 & 2.146 & PL & 2.00 & ..... & 25.58 & 32.14 & 57.73 & 43.94 & 44.04 & 44.30 \\
XMMC$\_$J095917.44+021514.9 & 170 & 9:59:17.44 & 2:15:14.91 & 142 & 0.935 & PL & 2.00 & ..... & 37.05 & 46.55 & 83.59 & 43.21 & 43.31 & 43.57 \\
XMMC$\_$J100128.19+021819.9 & 171 & 10:01:28.19 & 2:18:19.86 & 133 & 1.187 & PL & 2.00 & ..... & 34.80 & 43.72 & 78.52 & 43.44 & 43.54 & 43.80 \\
XMMC$\_$J095921.15+020030.8 & 196 & 9:59:21.15 & 2:00:30.83 & 154 & 1.486 & PL & 2.00 & ..... & 56.39 & 70.85 & 127.24 & 43.89 & 43.99 & 44.25 \\
XMMC$\_$J100047.93+014935.9 & 198 & 10:00:47.93 & 1:49:35.93 & 134 & 0.893 & PL & 2.00 & ..... & 46.55 & 58.49 & 105.04 & 43.26 & 43.36 & 43.62 \\
XMMC$\_$J095858.95+020138.7 & 199 & 9:58:58.95 & 2:01:38.72 & 268 & 2.454 & PL & 2.11$^{2.36}_{1.92}$ & ..... & 161.11 & 171.30 & 332.41 & 44.88 & 44.91 & 45.20 \\
XMMC$\_$J100105.90+015918.6 & 206 & 10:01:05.90 & 1:59:18.58 & 131 & 0.721 & APL & 2.00 & 21.69$^{22.19}_{20.99}$ & 29.60 & 53.14 & 82.74 & 43.00 & 43.10 & 43.35 \\
XMMC$\_$J100058.47+015206.4 & 216 & 10:00:58.47 & 1:52:06.40 & 216 & 2.029 & PL & 2.22$^{2.69}_{1.84}$ & ..... & 42.84 & 39.07 & 81.91 & 44.11 & 44.07 & 44.39 \\
XMMC$\_$J095956.08+014728.0 & 222 & 9:59:56.08 & 1:47:27.97 & 237 & 0.337 & PL & 2.28$^{2.57}_{2.01}$ & ..... & 73.53 & 61.26 & 134.79 & 42.57 & 42.49 & 42.83 \\
XMMC$\_$J100139.88+023132.8 & 236 & 10:01:39.88 & 2:31:32.77 & 110 & 1.444 & PL & 2.00 & ..... & 12.81 & 16.09 & 28.90 & 43.22 & 43.32 & 43.57 \\
XMMC$\_$J100046.86+014737.1 & 256 & 10:00:46.86 & 1:47:37.14 & 113 & 1.867 & APL & 2.00 & 20.71$^{21.83}_{20.42}$ & 55.75 & 70.92 & 126.67 & 44.22 & 44.32 & 44.57 \\
XMMC$\_$J100042.36+014535.7 & 265 & 10:00:42.36 & 1:45:35.66 & 101 & 1.161 & PL & 2.00 & ..... & 38.62 & 48.52 & 87.14 & 43.47 & 43.56 & 43.82 \\
XMMC$\_$J095910.00+022018.4 & 268 & 9:59:10.00 & 2:20:18.42 & 143 & 0.432 & APL & 2.00 & 21.32$^{21.60}_{20.89}$ & 34.78 & 56.34 & 91.12 & 42.49 & 42.59 & 42.84 \\
XMMC$\_$J100005.52+023057.4 & 274 & 10:00:05.52 & 2:30:57.40 & 112 & 0.677 & APL$+$po & 2.00 & 22.67$^{23.00}_{22.18}$ & 18.30 & 104.83 & 123.13 & 43.26 & 43.36 & 43.62 \\
XMMC$\_$J095929.40+022035.6 & 282 & 9:59:29.40 & 2:20:35.60 & 150 & 1.733 & PL & 2.00 & ..... & 18.24 & 22.92 & 41.16 & 43.57 & 43.67 & 43.92 \\
XMMC$\_$J095902.45+022510.6 & 288 & 9:59:02.45 & 2:25:10.61 & 202 & 1.105 & PL & 2.17$^{2.46}_{1.91}$ & ..... & 35.06 & 34.23 & 69.29 & 43.30 & 43.29 & 43.60 \\
XMMC$\_$J095927.04+015340.8 & 293 & 9:59:27.04 & 1:53:40.84 & 222 & 0.444 & APL & 1.51$^{2.10}_{1.21}$ & 21.91$^{22.13}_{21.70}$ & 136.15 & 666.98 & 803.13 & 43.28 & 43.69 & 43.83 \\
XMMC$\_$J100016.65+021352.1 & 298 & 10:00:16.65 & 2:13:52.11 & 100 & 1.867 & PL & 2.00 & ..... & 23.63 & 29.70 & 53.33 & 43.76 & 43.86 & 44.11 \\
XMMC$\_$J100049.94+015230.8 & 359 & 10:00:49.94 & 1:52:30.79 & 222 & 1.156 & PL & 1.54$^{1.85}_{1.28}$ & ..... & 35.99 & 89.17 & 125.16 & 43.35 & 43.74 & 43.89 \\
XMMC$\_$J100118.89+020729.0 & 391 & 10:01:18.89 & 2:07:28.98 & 110 & 1.774 & PL & 2.00 & ..... & 50.15 & 62.98 & 113.13 & 44.03 & 44.13 & 44.39 \\
XMMC$\_$J100006.35+023342.0 & 398 & 10:00:06.35 & 2:33:42.01 & 131 & 0.745 & APL & 2.00 & 21.64$^{21.98}_{21.19}$ & 40.43 & 69.95 & 110.38 & 43.15 & 43.25 & 43.51 \\
XMMC$\_$J095944.64+022626.2 & 416 & 9:59:44.64 & 2:26:26.22 & 102 & 0.992 & APL & 2.00 & 22.19$^{22.57}_{21.72}$ & 18.76 & 45.28 & 64.04 & 43.28 & 43.38 & 43.63 \\
XMMC$\_$J100223.07+014715.1 & 2013 & 10:02:23.07 & 1:47:15.07 & 686 & 1.243 & PL & 1.82$^{1.92}_{1.73}$ & ..... & 428.34 & 700.54 & 1128.90 & 44.58 & 44.80 & 45.00 \\
XMMC$\_$J095819.89+022903.8 & 2016 & 9:58:19.89 & 2:29:03.78 & 768 & 0.345 & PL & 2.19$^{2.31}_{2.07}$ & ..... & 385.27 & 373.31 & 758.58 & 43.21 & 43.19 & 43.50 \\
XMMC$\_$J100234.40+015011.5 & 2020 & 10:02:34.40 & 1:50:11.51 & 651 & 1.506 & PL & 2.25$^{2.37}_{2.14}$ & ..... & 226.06 & 196.73 & 422.80 & 44.51 & 44.45 & 44.78 \\
XMMC$\_$J100129.41+013633.7 & 2021 & 10:01:29.41 & 1:36:33.75 & 271 & 0.104 & APL & 1.23$^{1.39}_{1.01}$ & 22.38$^{22.51}_{22.24}$ & 163.24 & 4230.20 & 4393.50 & 42.39 & 43.12 & 43.19 \\
XMMC$\_$J100211.31+013707.2 & 2028 & 10:02:11.31 & 1:37:07.15 & 293 & 0.784 & APL+Fe & 2.55$^{2.79}_{2.29}$ & 21.83$^{21.95}_{21.69}$ & 113.12 & 204.74 & 317.86 & 43.75 & 43.50 & 43.94 \\
XMMC$\_$J100257.55+015405.6 & 2036 & 10:02:57.55 & 1:54:05.58 & 233 & 0.971 & PL & 1.89$^{2.07}_{1.71}$ & ..... & 284.95 & 423.49 & 708.44 & 44.14 & 44.31 & 44.54 \\
XMMC$\_$J100033.51+013812.6 & 2040 & 10:00:33.51 & 1:38:12.61 & 317 & 0.520 & PL & 2.28$^{2.47}_{2.10}$ & ..... & 263.04 & 212.05 & 475.09 & 43.40 & 43.32 & 43.67 \\
XMMC$\_$J100237.09+014648.3 & 2043 & 10:02:37.09 & 1:46:48.33 & 347 & 0.668 & APL+Fe & 1.56$^{1.92}_{1.31}$ & 21.77$^{21.96}_{21.52}$ & 172.69 & 608.59 & 781.28 & 43.55 & 43.93 & 44.08 \\
XMMC$\_$J100303.04+015209.2 & 2046 & 10:03:03.04 & 1:52:09.19 & 341 & 1.800 & PL & 2.23$^{2.43}_{2.04}$ & ..... & 132.62 & 119.47 & 252.09 & 44.47 & 44.43 & 44.75 \\
XMMC$\_$J100151.19+020032.8 & 2058 & 10:01:51.19 & 2:00:32.81 & 779 & 0.964 & PL & 2.02$^{2.13}_{1.91}$ & ..... & 285.44 & 348.97 & 634.41 & 44.13 & 44.22 & 44.48 \\
XMMC$\_$J100229.27+014528.2 & 2071 & 10:02:29.27 & 1:45:28.21 & 328 & 0.876 & PL & 1.58$^{1.73}_{1.44}$ & ..... & 177.16 & 414.17 & 591.34 & 43.72 & 44.09 & 44.24 \\
XMMC$\_$J100141.42+021031.8 & 2078 & 10:01:41.42 & 2:10:31.78 & 195 & 0.982 & APL & 1.93$^{2.20}_{1.59}$ & 20.96$^{21.50}_{20.42}$ & 135.04 & 197.89 & 332.93 & 43.85 & 44.00 & 44.23 \\
XMMC$\_$J100238.78+013938.2 & 2080 & 10:02:38.78 & 1:39:38.25 & 238 & 1.315 & PL & 1.87$^{2.02}_{1.73}$ & ..... & 127.19 & 193.62 & 320.82 & 44.07 & 44.26 & 44.48 \\
XMMC$\_$J100238.27+013747.8 & 2093 & 10:02:38.27 & 1:37:47.75 & 222 & 2.506 & PL & 1.97$^{2.16}_{1.79}$ & ..... & 131.93 & 172.73 & 304.66 & 44.82 & 44.94 & 45.18 \\
XMMC$\_$J100214.21+020620.0 & 2096 & 10:02:14.21 & 2:06:20.02 & 482 & 1.265 & PL & 1.64$^{1.75}_{1.54}$ & ..... & 131.85 & 282.01 & 413.85 & 44.03 & 44.36 & 44.53 \\
XMMC$\_$J100219.58+015536.9 & 2105 & 10:02:19.58 & 1:55:36.94 & 323 & 1.509 & PL & 2.19$^{2.43}_{1.97}$ & ..... & 74.17 & 70.26 & 144.43 & 44.03 & 44.01 & 44.32 \\
XMMC$\_$J100305.20+015157.0 & 2118 & 10:03:05.20 & 1:51:57.04 & 195 & 0.969 & APL & 2.14$^{2.61}_{1.75}$ & 20.99$^{21.58}_{20.42}$ & 165.82 & 180.69 & 346.51 & 43.93 & 43.94 & 44.24 \\
XMMC$\_$J095848.84+023442.3 & 2138 & 9:58:48.84 & 2:34:42.34 & 729 & 1.551 & PL & 2.01$^{2.11}_{1.90}$ & ..... & 121.90 & 151.84 & 273.74 & 44.28 & 44.37 & 44.63 \\
XMMC$\_$J100230.13+014810.0 & 2152 & 10:02:30.13 & 1:48:10.01 & 281 & 0.626 & PL & 2.21$^{2.57}_{1.90}$ & ..... & 94.64 & 86.97 & 181.61 & 43.19 & 43.16 & 43.48 \\
XMMC$\_$J100232.55+014009.5 & 2169 & 10:02:32.55 & 1:40:09.53 & 144 & 1.776 & PL & 2.00 & ..... & 72.41 & 90.98 & 163.39 & 44.19 & 44.29 & 44.55 \\
XMMC$\_$J100141.11+021259.9 & 2191 & 10:01:41.11 & 2:12:59.88 & 225 & 0.621 & PL & 2.27$^{2.53}_{2.03}$ & ..... & 87.36 & 74.02 & 161.38 & 43.01 & 42.94 & 43.28 \\
XMMC$\_$J100236.79+015948.5 & 2202 & 10:02:36.79 & 1:59:48.50 & 142 & 1.516 & PL & 2.00 & ..... & 65.21 & 79.78 & 144.99 & 43.98 & 44.07 & 44.33 \\
XMMC$\_$J100038.40+013708.4 & 2211 & 10:00:38.40 & 1:37:08.37 & 153 & 1.251 & PL & 2.00 & ..... & 65.00 & 81.68 & 146.68 & 43.77 & 43.87 & 44.13 \\
XMMC$\_$J100156.40+014811.0 & 2213 & 10:01:56.40 & 1:48:11.00 & 263 & 0.957 & APL & 2.02$^{2.53}_{1.64}$ & 20.87$^{21.55}_{20.42}$ & 71.18 & 91.34 & 162.52 & 43.54 & 43.63 & 43.89 \\
XMMC$\_$J100226.77+014052.1 & 2218 & 10:02:26.77 & 1:40:52.05 & 123 & 0.247 & PL & 2.00 & ..... & 55.24 & 69.41 & 124.65 & 42.01 & 42.11 & 42.36 \\
XMMC$\_$J100041.57+013658.7 & 2220 & 10:00:41.57 & 1:36:58.69 & 162 & 0.995 & PL & 2.00 & ..... & 73.60 & 92.48 & 166.08 & 43.58 & 43.68 & 43.93 \\
XMMC$\_$J100156.31+020942.9 & 2232 & 10:01:56.31 & 2:09:42.91 & 131 & 1.641 & PL & 2.00 & ..... & 43.15 & 54.21 & 97.36 & 43.88 & 43.98 & 44.24 \\
XMMC$\_$J100253.16+013457.8 & 2235 & 10:02:53.16 & 1:34:57.85 & 100 & 2.248 & PL & 2.00 & ..... & 65.71 & 81.11 & 146.82 & 44.40 & 44.49 & 44.75 \\
XMMC$\_$J095904.34+022552.8 & 2237 & 9:59:04.34 & 2:25:52.75 & 192 & 0.941 & APL & 1.78$^{2.41}_{1.40}$ & 22.74$^{22.96}_{22.55}$ & 38.88 & 285.86 & 324.74 & 43.91 & 44.15 & 44.34 \\
XMMC$\_$J100223.02+020639.5 & 2246 & 10:02:23.02 & 2:06:39.48 & 303 & 0.899 & PL & 1.95$^{2.15}_{1.76}$ & ..... & 69.93 & 94.68 & 164.61 & 43.41 & 43.54 & 43.78 \\
XMMC$\_$J100243.88+020501.6 & 2261 & 10:02:43.88 & 2:05:01.59 & 206 & 1.234 & PL & 1.97$^{2.27}_{1.70}$ & ..... & 72.98 & 95.93 & 168.91 & 43.81 & 43.93 & 44.17 \\
XMMC$\_$J100208.53+014553.7 & 2276 & 10:02:08.53 & 1:45:53.65 & 111 & 2.215 & PL & 2.00 & ..... & 31.49 & 39.56 & 71.05 & 44.07 & 44.17 & 44.42 \\
XMMC$\_$J100158.05+014621.7 & 2289 & 10:01:58.05 & 1:46:21.74 & 122 & 0.831 & APL & 2.00 & 22.77$^{22.97}_{22.54}$ & 32.93 & 230.29 & 263.22 & 43.83 & 43.93 & 44.19 \\
XMMC$\_$J100130.33+014305.0 & 2299 & 10:01:30.33 & 1:43:04.97 & 110 & 1.571 & PL & 2.00 & ..... & 76.83 & 96.53 & 173.35 & 44.09 & 44.19 & 44.44 \\
XMMC$\_$J100143.54+015606.2 & 2361 & 10:01:43.54 & 1:56:06.18 & 195 & 2.181 & PL & 1.94$^{2.24}_{1.68}$ & ..... & 93.06 & 127.11 & 220.17 & 44.52 & 44.66 & 44.90 \\
XMMC$\_$J100240.34+020146.4 & 2370 & 10:02:40.34 & 2:01:46.37 & 132 & 0.638 & APL & 2.00 & 22.16$^{22.69}_{21.76}$ & 34.05 & 103.02 & 137.07 & 43.17 & 43.27 & 43.52 \\
XMMC$\_$J100141.54+020051.4 & 2557 & 10:01:41.54 & 2:00:51.44 & 120 & 2.277 & PL & 2.00 & ..... & 158.62 & 199.31 & 357.93 & 44.80 & 44.90 & 45.15 \\
XMMC$\_$J100142.26+020358.5 & 2608 & 10:01:42.26 & 2:03:58.49 & 131 & 0.125 & PEXRAV+Fe & 2.00 & $>$24.18 & 28.99 & 511.11 & 63.12 & 41.07 & 42.32 & 43.87 \\
XMMC$\_$J100136.21+015442.5 & 2703 & 10:01:36.21 & 1:54:42.45 & 151 & 2.281 & PL & 2.00 & ..... & 209.26 & 262.93 & 472.19 & 44.92 & 45.02 & 45.27 \\
\enddata
\tablenotetext{a}{IAU name.}
\tablenotetext{b}{Internal reference number.}
\tablenotetext{c}{X-ray coordinates.}
\tablenotetext{d}{Net {\it pn} counts in the [0.3-10] keV energy band.}
\tablenotetext{e}{Spectroscopic redshift of the most likely optical counterpart (for details see \citealt{brusa06}).}
\tablenotetext{f}{Best fit model as discussed in \S4.}
\tablenotetext{g}{X-ray fluxes from the spectral fit in units of $10^{-16}$ erg cm$^{-2}$ s$^{-1}$ in the [0.5-2], [2-10] and [0.5-10] keV rest-frame energy bands respectively}
\tablenotetext{h}{ ${\rm Logarithm}$ of the X-ray luminosities corrected for absorption in the [0.5-2], [2-10] and [0.5-10] keV rest-frame energy bands respectively.}
\tablenotetext{i}{A powerlaw model is not a good representation of this source(see \S4.1).}
\end{deluxetable}
\clearpage
\end{landscape}

 \begin{deluxetable}{lccc}
 \tabletypesize{\scriptsize}
 \tablecaption{Comparison between optical and X-ray classifications
 \label{opt_xray}}
 \tablewidth{0pt}
 \tablehead{
  &  \colhead{Broad Line AGN} & \colhead{Narrow Line AGN} & \colhead{Galaxy} } 
 \startdata
X-ray unabsorbed AGN & 78 & 19 & 5 \\ 
X-ray absorbed AGN & 8 & 13 & 11 \\ 
X-ray galaxy & 0 & 0 & 0 \\ 
 \enddata
  \end{deluxetable}

 
 \end{document}